\documentclass[a4paper,fleqn,usenatbib]{mnras}

\usepackage{newtxtext,newtxmath}

\usepackage[T1]{fontenc}
\usepackage{ae,aecompl}

\usepackage{graphicx}	
\usepackage{amsmath}	
\usepackage{amssymb}	

\usepackage{booktabs}
\usepackage[flushleft]{threeparttable}
\usepackage{tablefootnote}

\title[MASSIVE -- X. Kinemetry]{The MASSIVE Survey -- X. Misalignment between Kinematic and Photometric Axes and Intrinsic Shapes of Massive Early-Type Galaxies}

\author[Ene et al.]{
Irina Ene,$^{1, 2}$\thanks{E-mail: irina.ene@berkeley.edu}
Chung-Pei Ma,$^{1, 2}$ Melanie Veale,$^{1, 2}$ Jenny E. Greene,$^{3}$ Jens Thomas,$^{4}$
\newauthor John P. Blakeslee,$^{5}$ Caroline Foster,$^{6,7}$ Jonelle L. Walsh,$^{8}$  Jennifer Ito,$^{9}$ 
\newauthor and Andy D. Goulding$^{3}$
\\
$^{1}$Department of Astronomy, University of California, Berkeley, CA 94720, USA\\
$^{2}$Department of Physics, University of California, Berkeley, CA 94720, USA\\
$^{3}$Department of Astrophysical Sciences, Princeton University, Princeton, NJ 08544, USA\\
$^{4}$Max Plank-Institute for Extraterrestrial Physics, Giessenbachstr. 1, D-85741 Garching, Germany\\
$^{5}$Dominion Astrophysical Observatory, NRC Herzberg Astronomy and Astrophysics, Victoria, BC V9E 2E7, Canada\\
$^{6}$Sydney Institute for Astronomy, School of Physics, A28, The University of Sydney, NSW, 2006, Australia\\
$^{7}$ARC Centre of Excellence for All Sky Astrophysics in 3 Dimensions (ASTRO 3D)\\
$^{8}$George P. and Cynthia Woods Mitchell Institute for Fundamental Physics and Astronomy, and Department of Physics and Astronomy, \\
Texas A\&M University, College Station, TX 77843, USA\\
$^{9}$Department of Physics, University of California, San Diego, CA 92093, USA\\
}

\date{Accepted XXX. Received YYY; in original form ZZZ}

\pubyear{2018}

\begin{document}
\label{firstpage}
\pagerange{\pageref{firstpage}--\pageref{lastpage}}
\maketitle

\begin{abstract}
We use spatially resolved two-dimensional stellar velocity maps over a $107''\times 107''$ field of view to investigate the kinematic features of 90 early-type galaxies above stellar mass $10^{11.5}M_\odot$ in the MASSIVE survey. We measure the misalignment angle $\Psi$ between the kinematic and photometric axes and identify local features such as velocity twists and kinematically distinct components.
We find 46\% of the sample to be well aligned ($\Psi < 15\degr$), 33\% misaligned, and 21\% without detectable rotation (non-rotators).
Only 24\% of the sample are fast rotators, the majority of which (91\%) are aligned, whereas 57\% of the slow rotators are misaligned with a nearly flat distribution of $\Psi$ from $15\degr$ to $90\degr$.
11 galaxies have $\Psi \gtrsim 60\degr$ and thus exhibit minor-axis (``prolate'') rotation  in which the rotation is preferentially around the photometric major axis.
Kinematic misalignments occur more frequently for lower galaxy spin or denser galaxy environments.
Using the observed misalignment and ellipticity distributions, we infer the intrinsic shape distribution of our sample and find that MASSIVE slow rotators are consistent with being mildly triaxial, with mean axis ratios of $b/a=0.88$ and $c/a=0.65$.
In terms of local kinematic features, 51\% of the sample exhibit kinematic twists of larger than $20\degr$, and 2 galaxies have kinematically distinct components.
The frequency of misalignment and the broad distribution of $\Psi$ reported here suggest that the most massive early-type galaxies are 
mildly triaxial, and that formation processes resulting in kinematically misaligned slow rotators such as gas-poor mergers occur frequently in this mass range.
\end{abstract}




\section{Introduction}

Understanding the distribution of the intrinsic shapes of early-type galaxies plays a key role in the study of galaxy formation and evolution.
The two-dimensional surface brightness profile of an individual galaxy alone can not uniquely determine its three-dimensional intrinsic shape. Comparing stellar kinematics along the apparent major and minor axes can provide additional information about the triaxiality of an early-type galaxy \citep{Binney1985}.

Early spectroscopic observations along both the major and minor axes of elliptical galaxies found a varying degree of rotation along each axis (e.g., \citealt{SchechterGunn1979,DaviesBirkinshaw1986,DaviesBirkinshaw1988,Wagneretal1988,FranxIllingworth1988,JedrzejewskiSchechter1989,Franxetal1991}),
lending support to the notion that elliptical galaxies can have intrinsically triaxial shapes \citep{binney1978}.
An in-depth study by \citet{Franxetal1991} found that the observed apparent ellipticity and kinematic misalignment distributions of 38 elliptical galaxies can be explained by various intrinsic shape distributions, including all nearly oblate, a mixture of oblate and prolate, and all triaxial.

More recently, several surveys have taken advantage of integral-field spectroscopy (IFS) observations to obtain detailed two-dimensional maps of stellar kinematics, e.g., SAURON \citep{Emsellemetal2004}, ATLAS$^{\text{3D}}$ \citep{Cappellarietal2011a}, SAMI \citep{Croometal2012}, CALIFA \citep{Sanchezetal2012}, and MaNGA \citep{Bundyetal2015} (see \citealt{Cappellari2016} for a review). In addition, a number of IFS studies of early-type galaxies targeted specifically brightest cluster galaxies (BCGs), e.g., \citet{Broughetal2011, Jimmyetal2013}, and
a few IFS (or multi-slit) studies had wide enough sky coverage to reach $\sim 2-4$ effective radii, e.g., SLUGGS \citep{Brodieetal2014} and \citet{raskuttietal2014}.
A major goal of these studies was to investigate the properties and statistics of fast rotators (FR) and slow rotators (SR).  Some of these studies further investigated the misalignment between kinematic and photometric axes and local velocity features in early-type galaxies.
This was achieved by using the kinemetry method \citep{Krajnovicetal2006}, a generalization of isophotal analysis of light distributions to the analysis of the spatial properties of velocity features.
It was found that, in general, fast rotators have well aligned photometric and kinematic axes and show relatively featureless velocity maps or small kinematic twists, while slow rotators tend to be misaligned, show large kinematic twists or complex kinematic features,
and are mildly triaxial (e.g., \citealt{emsellemetal2007, krajnovicetal2011, Weijmansetal2014}).
Measuring the kinematics of the outer parts of elliptical galaxies further than a few effective radii is difficult to achieve using absorption spectroscopy due to the decrease in surface brightness at large radii.
Instead, one must use discrete tracers such as planetary nebulae, which have been shown to be good tracers of stellar kinematics in the halos of early-type galaxies \citep{Coccatoetal2009, Cortesietal2013}.
Results from planetary nebulae derived velocity fields \citep{Pulsonietal2017} show that the halos of elliptical galaxies exhibit more diverse kinematic properties, such as misaligned halos in fast rotators and decoupled halos in slow rotators.

Aside from the BCG studies, the early-type galaxies targeted in the aforementioned spectroscopic surveys were predominantly fast-rotating, lower-mass S0 or elliptical galaxies. For example, ATLAS$^{\text{3D}}$ studied 260 early-type galaxies with stellar mass $M_* \gtrsim 6 \times 10^9 M_\odot$ in a volume up to 42 Mpc. They found 86\% of the galaxies to be fast rotators, and 90\% of the galaxies to show alignment between the kinematic and photometric axes
\citep{emsellemetal2011}.
In the SAMI pilot sample of 106 galaxies,
80 are early-types, and only $20\%$ of them are classified as slow rotators \citep{Fogartyetal2015}. They found 83\% of their fast rotators to be aligned, while only 38\% of the slow rotators are aligned. A study of 315 early-type galaxies in galaxy clusters in the SAMI survey found only 14\% to be slow rotators \citep{Broughetal2017}.

We designed the MASSIVE survey \citep{Maetal2014} to
sample the high-mass regime of early-type galaxies that was little explored in previous IFS surveys.
MASSIVE is volume-limited, surveying the $\sim 100$ most massive galaxies located up to a distance of 108 Mpc in the northern sky. This survey is complete to an absolute $K$-band magnitude of $M_K = -25.3$ mag, or
stellar mass $M_* \ga 10^{11.5} M_\odot$.
In a series of three papers \citep{Vealeetal2017a, Vealeetal2017b, Vealeetal2018}, we investigated a wide range of stellar kinematic properties of MASSIVE galaxies
using spatially-resolved IFS measurements of stellar velocities, velocity dispersions, and higher velocity moments over a $107''\times 107''$ field of view (FOV).
We found the fraction of slow rotators to rise rapidly with increasing $M_*$ in our mass range, reaching $\sim 90$\% at $M_* \sim 10^{12} M_\odot$ (Paper V; \citealt{Vealeetal2017a}).
Paper VII \citep{Vealeetal2017b} examined the relationships between galaxy spin, stellar mass, and environment in a combined sample of 370 MASSIVE and ATLAS$^{\text{3D}}$ survey galaxies.
We found that the apparent kinematic morphology-density relation for local early-type galaxies is primarily driven by
galaxy mass rather than galaxy environment.
The radial profiles of velocity dispersions
for 90 MASSIVE galaxies out to projected radii as large as 30 kpc were presented in Paper VIII \citep{Vealeetal2018}.
We found the fraction of galaxies with rising outer profiles to increase with $M_*$ and in denser galaxy environments, and this trend
is likely to be caused by variations in the total mass profiles rather than in the velocity anisotropy alone.
Additionally, Paper VI \citep{Pandyaetal2017} analysed the morphology and kinematics of warm ionized gas detected in MASSIVE galaxies. We found kinematic misalignment between stars and gas to be more prevalent in slow rotators (2/2) than in fast rotators (1/4), pointing to external means of acquiring gas for massive early-type galaxies.

This paper is devoted to the stellar velocity properties of MASSIVE galaxies.
Here we make use of the two-dimensional nature of IFS data to analyse the stellar velocity features of 90 MASSIVE galaxies.
We use the kinemetry method \citep{Krajnovicetal2006} to determine the kinematic features and misalignment angle of each galaxy.
We investigate the correlations between kinematic misalignment and galaxy morphology and environment for fast and slow rotating galaxies.

The paper is organized as follows. 
The galaxy sample and data reduction procedure are described in Section~\ref{sec:observations}.
Section~\ref{sec:kinematic_analysis} describes how the kinemetry method is used to measure the global and local velocity properties.  
In Section~\ref{sec:global_kinematic_alignment} we investigate the distribution of the misalignment between the photometric and kinematic axes and the correlations between alignment and galaxy morphological parameters such as ellipticity, stellar mass, or environmental parameters. 
In Section~\ref{sec:intrinsic_shapes} we invert the observed kinematic misalignment and ellipticity distributions to determine the intrinsic shape distribution of our galaxy sample.
Section~\ref{sec:local_kinematic_features} presents the local kinematic features we identified on the velocity maps and discusses the more interesting cases. 
We enumerate the main conclusions and discuss the implications of our findings in Section~\ref{sec:conclusions}.


\section{Observations and Data}
\label{sec:observations}

\begin{table*}
  \centering
  \caption{Key kinematic properties of MASSIVE galaxies.}
  \label{tab:main}
 \begin{tabular}{llccrcrrcc}
\toprule
\midrule
\multicolumn{1}{c}{Name} & $\varepsilon$ & $\lambda_e$ & $\lambda_e$ & PA$_{\text{phot}}$ & \multicolumn{1}{c}{PA$_{\text{kin}}$}  & $\Psi$ &  $k_1^{\text{max}}$ & Alignment/ & Features \\
& & [folded] & [unfolded] & $\lbrack$deg$\rbrack$ & \multicolumn{1}{c}{$\lbrack$deg$\rbrack$} &  $\lbrack$deg$\rbrack$ & $\lbrack$km/s$\rbrack$ & Rotator & \\
\multicolumn{1}{c}{(1)} & (2) & (3) & (4) & (5) & \multicolumn{1}{c}{(6)} & (7) & (8) & (9) & (10)\\
\midrule
IC0310 & 0.06 & 0.085 & 0.091 & 32.6 & 192.0 $\pm$ 13.8 & 20.6 & 30.5 & M-SR$^4$ & Kinematic Twist \\
NGC0057 & 0.17 & 0.022 & 0.028 & 41.1 & - & - & 9.6 & None & - \\
NGC0080 & 0.09 & 0.039 & 0.045 & 9.6 & 28.5 $\pm$ 28.5 & 18.9 & 13.6 & A-SR & Regular Rotation$^*$ \\
NGC0315 & 0.28 & 0.062 & 0.063 & 42.4 & 222.0 $\pm$ 7.2 & 0.4 & 44.1 & A-SR & Regular Rotation \\
NGC0383 & 0.14 & 0.252 & 0.247 & 141.2 & 140.5 $\pm$ 2.8 & 0.7 & 103.8 & A-FR & Regular Rotation \\
NGC0410 & 0.25 & 0.034 & 0.048 & 34.9 & 161.0 $\pm$ 18.5 & 53.9 & 18.9 & M-SR & Kinematic Twist$^*$ \\
NGC0499 & 0.35 & 0.060 & 0.066 & 73.8 & 273.5 $\pm$ 8.0 & 19.7 & 46.2 & M-SR & Regular Rotation \\
NGC0507 & 0.09 & 0.049 & 0.082 & 21.9 & 183.5 $\pm$ 4.0 & 18.4 & 50.4 & M-SR & KDC \\
NGC0533 & 0.26 & 0.034 & 0.051 & 51.2 & 6.0 $\pm$ 24.8 & 45.2 & 19.8 & M-SR & Kinematic Twist \\
NGC0545 & 0.28 & 0.129 & 0.081 & 59.7 & - & - & 10.9 & None & - \\
NGC0547 & 0.14 & 0.056 & 0.081 & 94.1 & - & - & 29.6 & None & - \\
NGC0665 & 0.24 & 0.402 & 0.395 & 114.8 & 116.0 $\pm$ 3.2 & 1.2 & 126.0 & A-FR & Regular Rotation \\
NGC0708 & 0.40$\dagger$ & 0.036 & 0.043 & 45.0$\dagger$ & 126.0 $\pm$ 17.5 & 81.0 & 31.8 & M-SR & Regular Rotation \\
NGC0741 & 0.17 & 0.037 & 0.050 & 86.7 & - & - & 12.4 & None & - \\
NGC0777 & 0.17 & 0.046 & 0.060 & 148.4 & 8.0 $\pm$ 10.0 & 39.6 & 41.0 & M-SR & Kinematic Twist \\
NGC0890 & 0.38$\dagger$ & 0.101 & 0.136 & 55.0$\dagger$ & 100.5 $\pm$ 9.2 & 45.5 & 45.7 & M-SR & Kinematic Twist \\
NGC0910 & 0.16$\dagger$ & 0.039 & 0.050 & 30.0$\dagger$ & 59.0 $\pm$ 27.8 & 29.0 & 14.5 & A-SR & Regular Rotation \\
NGC0997 & 0.13 & 0.243 & 0.240 & 14.7 & 219.5 $\pm$ 6.2 & 24.8 & 93.5 & M-FR & Kinematic Twist \\
NGC1016 & 0.06 & 0.033 & 0.040 & 40.5 & 261.5 $\pm$ 19.5 & 41.0 & 29.9 & M-SR & Kinematic Twist$^*$ \\
NGC1060 & 0.24 & 0.023 & 0.048 & 74.0 & 342.0 $\pm$ 13.5 & 88.0 & 14.9 & M-SR & Kinematic Twist$^*$ \\
NGC1129 & 0.15$^1$ & 0.120 & 0.124 & 46.2 & 179.0 $\pm$ 6.0 & 47.2 & 66.2 & M-SR & Kinematic Twist \\
NGC1132 & 0.37 & 0.061 & 0.082 & 141.3 & 122.0 $\pm$ 18.2 & 19.3 & 26.3 & A-SR & Kinematic Twist$^*$ \\
NGC1167 & 0.17 & 0.427 & 0.395 & 71.4 & 254.0 $\pm$ 3.8 & 2.6 & 110.0 & A-FR & Regular Rotation \\
NGC1226 & 0.18$\dagger$ & 0.033 & 0.030 & 90.0$\dagger$ & - & - & 22.5 & None & - \\
NGC1272 & 0.07 & 0.023 & 0.037 & 160.3 & - & - & 11.4 & None & - \\
NGC1453 & 0.14$\dagger$ & 0.201 & 0.204 & 35.0$\dagger$ & 35.0 $\pm$ 3.2 & 0.0 & 91.8 & A-FR & Regular Rotation \\
NGC1497 & 0.40$\dagger$ & 0.474 & 0.466 & 60.0$\dagger$ & 54.5 $\pm$ 3.5 & 5.5 & 161.1 & A-FR & Regular Rotation \\
NGC1573 & 0.34$\dagger$ & 0.040 & 0.056 & 35.0$\dagger$ & 189.5 $\pm$ 19.2 & 25.5 & 25.5 & M-SR & Kinematic Twist \\
NGC1600 & 0.26$\dagger$ & 0.026 & 0.035 & 10.0$\dagger$ & - & - & 21.8 & None & - \\
NGC1684 & 0.24$\dagger$ & 0.122 & 0.133 & 95.0$\dagger$ & 223.0 $\pm$ 1.0 & 52.0 & 41.8 & M-SR & Kinematic Twist$^*$ \\
NGC1700 & 0.28$\dagger$ & 0.195 & 0.198 & 90.0$\dagger$ & 268.0 $\pm$ 1.8 & 2.0 & 114.9 & A-FR & Regular Rotation \\
NGC2208 & 0.32$\dagger$ & 0.062 & 0.049 & 95.0$\dagger$ & - & - & 16.9 & None & - \\
NGC2256 & 0.20$\dagger$ & 0.048 & 0.056 & 75.0$\dagger$ & - & - & 17.0 & None & - \\
NGC2258 & 0.24$\dagger$ & 0.036 & 0.071 & 135.0$\dagger$ & 71.0 $\pm$ 17.2 & 64.0 & 30.6 & M-SR & Regular Rotation \\
NGC2274 & 0.10$\dagger$ & 0.067 & 0.073 & 145.0$\dagger$ & 287.5 $\pm$ 25.8 & 37.5 & 25.9 & M-SR & Kinematic Twist \\
NGC2320 & 0.30$\dagger$ & 0.233 & 0.214 & 140.0$\dagger$ & 147.0 $\pm$ 4.0 & 7.0 & 112.7 & A-FR & Regular Rotation \\
NGC2332 & 0.34$\dagger$ & 0.037 & 0.034 & 65.0$\dagger$ & - & - & 14.9 & None & - \\
NGC2340 & 0.44$\dagger$ & 0.029 & 0.032 & 80.0$\dagger$ & - & - & 12.4 & None & - \\
NGC2418 & 0.20 & 0.241 & 0.242 & 63.9 & 77.5 $\pm$ 3.2 & 13.6 & 100.4 & A-FR & Regular Rotation \\
NGC2513 & 0.20 & 0.095 & 0.101 & 166.8 & 348.0 $\pm$ 9.0 & 1.2 & 44.0 & A-SR & Regular Rotation \\
NGC2672 & 0.14 & 0.095 & 0.106 & 126.0 & 167.5 $\pm$ 5.2 & 41.5 & 44.8 & M-SR & Kinematic Twist \\
NGC2693 & 0.25 & 0.295 & 0.294 & 166.5 & 169.0 $\pm$ 2.2 & 2.5 & 144.2 & A-FR & Regular Rotation \\
NGC2783 & 0.39 & 0.042 & 0.060 & 165.2 & 269.5 $\pm$ 11.0 & 75.7 & 29.4 & M-SR & Kinematic Twist$^*$ \\
NGC2832 & 0.31 & 0.071 & 0.084 & 156.2 & 253.0 $\pm$ 31.0 & 83.2 & 32.2 & M-SR & Kinematic Twist \\
NGC2892 & 0.06 & 0.070 & 0.070 & 138.4 & - & - & 18.3 & None & - \\
NGC3158 & 0.18 & 0.255 & 0.258 & 152.6 & 330.0 $\pm$ 2.2 & 2.6 & 127.0 & A-FR & Regular Rotation \\
NGC3209 & 0.27 & 0.039 & 0.051 & 82.2 & 323.0 $\pm$ 27.0 & 60.8 & 19.0 & M-SR & Regular Rotation \\
NGC3462 & 0.26 & 0.085 & 0.106 & 55.5 & 109.5 $\pm$ 11.2 & 54.0 & 51.5 & M-SR & Kinematic Twist \\
NGC3562 & 0.16$\dagger$ & 0.038 & 0.051 & 160.0$\dagger$ & 354.0 $\pm$ 27.0 & 14.0 & 26.3 & A-SR & Kinematic Twist \\
NGC3615 & 0.38 & 0.399 & 0.399 & 44.2 & 41.5 $\pm$ 3.2 & 2.7 & 136.7 & A-FR & Kinematic Twist \\
NGC3805 & 0.36 & 0.496 & 0.475 & 64.6 & 61.5 $\pm$ 1.5 & 3.1 & 184.9 & A-FR & Kinematic Twist \\
NGC3816 & 0.31 & 0.105 & 0.144 & 70.0 & 255.0 $\pm$ 12.8 & 5.0 & 35.1 & A-SR & Regular Rotation \\
NGC3842 & 0.22 & 0.038 & 0.040 & 1.6 & 185.0 $\pm$ 30.2 & 3.4 & 11.4 & A-SR & Kinematic Twist$^*$ \\
NGC3862 & 0.06 & 0.045 & 0.055 & 148.2 & - & - & 28.4 & None & - \\
NGC3937 & 0.20 & 0.074 & 0.060 & 27.3 & 200.0 $\pm$ 15.5 & 7.3 & 32.4 & A-SR & Kinematic Twist \\
NGC4073 & 0.32 & 0.023 & 0.038 & 101.3 & 339.5 $\pm$ 20.5 & 58.2 & 20.1 & M-SR & Kinematic Twist$^*$ \\
NGC4472 & 0.17$^1$ & 0.197 & 0.197 & 155.0 & 169.0 $\pm$ 5.5$^2$ & 14.0 & 58.9$^3$  & A-FR & - \\
NGC4555 & 0.20 & 0.120 & 0.121 & 117.7 & 119.5 $\pm$ 7.0 & 1.8 & 60.2 & A-SR & Kinematic Twist \\
NGC4816 & 0.20 & 0.069 & 0.073 & 85.6 & 239.0 $\pm$ 18.8 & 26.6 & 23.7 & M-SR & Kinematic Twist \\
NGC4839 & 0.35 & 0.045 & 0.045 & 65.0 & - & - & 19.0 & None & - \\
NGC4874 & 0.09 & 0.070 & 0.072 & 40.6 & 334.5 $\pm$ 5.5 & 66.1 & 40.1 & M-SR & Kinematic Twist \\
\midrule
\end{tabular}
\end{table*}
\begin{table*}
\centering
\label{table:maincontinued}
\contcaption{}
\begin{threeparttable}
\begin{tabular}{llccrcrrcc}
\toprule
\midrule
\multicolumn{1}{c}{Name} & $\varepsilon$ & $\lambda_e$ & $\lambda_e$ & PA$_{\text{phot}}$ & \multicolumn{1}{c}{PA$_{\text{kin}}$}  & $\Psi$ &  $k_1^{\text{max}}$ & Alignment/ & Features \\
& & [folded] & [unfolded] & $\lbrack$deg$\rbrack$ & \multicolumn{1}{c}{$\lbrack$deg$\rbrack$} &  $\lbrack$deg$\rbrack$ & $\lbrack$km/s$\rbrack$ & Rotator & \\
\multicolumn{1}{c}{(1)} & (2) & (3) & (4) & (5) & \multicolumn{1}{c}{(6)} & (7) & (8) & (9) & (10)\\
\midrule
NGC4889 & 0.36 & 0.032 & 0.033 & 80.3 & - & - & 24.2 & None & - \\
NGC4914 & 0.39 & 0.054 & 0.064 & 155.1 & 350.0 $\pm$ 13.2 & 14.9 & 23.3 & A-SR & Regular Rotation \\
NGC5129 & 0.37 & 0.402 & 0.389 & 5.6 & 3.0 $\pm$ 2.5 & 2.6 & 134.7 & A-FR & Regular Rotation \\
NGC5208 & 0.63 & 0.615 & 0.622 & 162.9 & 343.0 $\pm$ 1.5 & 0.1 & 243.2 & A-FR & Regular Rotation \\
NGC5322 & 0.33 & 0.054 & 0.054 & 92.0 & 267.5 $\pm$ 7.8 & 4.5 & 40.3 & A-SR & KDC \\
NGC5353 & 0.56 & 0.551 & 0.550 & 138.9 & 323.0 $\pm$ 0.5 & 4.1 & 285.0 & A-FR & Regular Rotation \\
NGC5490 & 0.20 & 0.138 & 0.136 & 3.4 & 183.0 $\pm$ 4.8 & 0.4 & 67.7 & A-SR & Regular Rotation \\
NGC5557 & 0.17 & 0.035 & 0.035 & 93.7 & 202.0 $\pm$ 16.2 & 71.7 & 27.4 & M-SR & Kinematic Twist$^*$ \\
NGC6223 & 0.20$\dagger$ & 0.315 & 0.320 & 85.0$\dagger$ & 288.0 $\pm$ 2.2 & 23.0 & 135.6 & M-FR & Kinematic Twist \\
NGC6375 & 0.10$\dagger$ & 0.240 & 0.232 & 145.0$\dagger$ & 322.0 $\pm$ 7.0 & 3.0 & 66.2 & A-FR & Regular Rotation \\
NGC6482 & 0.36$\dagger$ & 0.137 & 0.138 & 65.0$\dagger$ & 63.5 $\pm$ 2.0 & 1.5 & 118.4 & A-SR & Regular Rotation \\
NGC6575 & 0.28$\dagger$ & 0.124 & 0.136 & 65.0$\dagger$ & 65.0 $\pm$ 7.5 & 0.0 & 48.6 & A-SR & Kinematic Twist \\
NGC7052 & 0.50$\dagger$ & 0.148 & 0.147 & 65.0$\dagger$ & 61.5 $\pm$ 2.2 & 3.5 & 71.1 & A-SR & Regular Rotation$^*$ \\
NGC7242 & 0.28$\dagger$ & 0.037 & 0.043 & 40.0$\dagger$ & - & - & 17.1 & None & - \\
NGC7265 & 0.22$\dagger$ & 0.039 & 0.079 & 165.0$\dagger$ & 259.5 $\pm$ 9.5 & 85.5 & 30.5 & M-SR & Kinematic Twist \\
NGC7274 & 0.06$\dagger$ & 0.090 & 0.081 & 160.0$\dagger$ & 253.5 $\pm$ 12.5 & 86.5 & 31.9 & M-SR$^4$ & Kinematic Twist \\
NGC7386 & 0.28 & 0.071 & 0.090 & 140.9 & 111.0 $\pm$ 19.5 & 29.9 & 29.2 & M-SR & Kinematic Twist \\
NGC7426 & 0.34$\dagger$ & 0.563 & 0.525 & 70.0$\dagger$ & 251.0 $\pm$ 1.2 & 1.0 & 216.3 & A-FR & Regular Rotation \\
NGC7436 & 0.12 & 0.085 & 0.082 & 13.1 & 197.5 $\pm$ 10.2 & 4.4 & 47.6 & A-SR & Regular Rotation \\
NGC7550 & 0.07 & 0.038 & 0.040 & 155.2 & - & - & 14.9 & None & - \\
NGC7556 & 0.25 & 0.049 & 0.052 & 113.8 & 294.0 $\pm$ 30.8 & 0.2 & 54.9 & A-SR & Kinematic Twist$^*$ \\
NGC7618 & 0.28$\dagger$ & 0.247 & 0.239 & 0.0$\dagger$ & 178.0 $\pm$ 4.5 & 2.0 & 106.5 & A-FR & Regular Rotation \\
NGC7619 & 0.23 & 0.119 & 0.125 & 36.1 & 215.5 $\pm$ 2.2 & 0.6 & 53.3 & A-SR & Regular Rotation \\
NGC7626 & 0.14 & 0.034 & 0.040 & 10.5 & 2.5 $\pm$ 40.0 & 8.0 & 18.4 & A-SR & Kinematic Twist$^*$ \\
UGC01332 & 0.30$\dagger$ & 0.037 & 0.056 & 80.0$\dagger$ & - & - & 16.7 & None & - \\
UGC02783 & 0.11 & 0.068 & 0.082 & 113.9 & 9.0 $\pm$ 10.2 & 75.1 & 47.9 & M-SR & Kinematic Twist \\
UGC03683 & 0.26$\dagger$ & 0.090 & 0.102 & 45.0$\dagger$ & 49.5 $\pm$ 6.2 & 4.5 & 32.3 & A-SR & Kinematic Twist \\
UGC03894 & 0.10 & 0.122 & 0.130 & 35.2 & 217.0 $\pm$ 8.5 & 1.8 & 52.9 & A-FR & Regular Rotation \\
UGC10918 & 0.14$\dagger$ & 0.042 & 0.022 & 5.0$\dagger$ & - & - & 29.7 & None & - \\
\midrule
\end{tabular}

\begin{tablenotes}
    \item Column (1): Galaxy name.
    \item Column (2): Ellipticity, taken from NSA where available, otherwise from 2MASS (denoted by $\dagger$). $^1$ $\varepsilon$ is taken from our CFHT data for NGC 1129 and from \citet{emsellemetal2011} for NGC 4472.
    \item Column (3): Spin parameter at the effective radius, using the folded binning scheme of \citet{Vealeetal2017a}.
    \item Column (4): Spin parameter at the effective radius, using the unfolded binning scheme of this work.
    \item Column (5): Photometric position angle, taken from NSA where available, otherwise from 2MASS (denoted by $\dagger$).
    \item Column (6): Global kinematic position angle with 1-sigma errors. See Section~\ref{sub:pakin_psi} for definitions. $^2$ PA$_{\rm kin}$ is taken from \citet{krajnovicetal2011} for NGC~4472.
    \item Column (7): Kinematic misalignment angle. See Section~\ref{sub:pakin_psi} for definitions.
    \item Column (8): Maximum value of the kinemetry coefficient $k_1(R)$. See Section~\ref{sub:local_kinematic_features} for definitions. $^3$ $k_1^\text{max}$ is taken from \citet{krajnovicetal2011} for NGC~4472.
    \item Column (9): Kinematic misalignment category: aligned ('A') or misaligned ('M') and fast rotator ('FR') or slow rotator ('SR') classification. The non-rotating slow rotators are denoted by 'None'. $^4$ These galaxies would be fast rotators by using the ATLAS$^{\rm 3D}$ $\lambda_e$ -- $\varepsilon$ criterion. Based on the slow rotation seen on the velocity maps, we classify these galaxies as slow rotators.
    \item Column (10): Local kinematic features. See Section~\ref{sub:radial_profiles} for definitions. $^*$ Due to limited spatial coverage or poor S/N, these galaxies couldn't be classified using the criteria in Section~\ref{sub:radial_profiles}. Instead, we manually assign the closest matching kinematic feature.
\end{tablenotes}

\end{threeparttable}

\end{table*}

\subsection{Galaxy sample and IFS data}
\label{sub:ifsdata}

We study the velocity features of 90 early-type galaxies in the MASSIVE survey \citep{Maetal2014}.
The list of galaxies and key galaxy parameters are given in Table~\ref{tab:main}.
The MASSIVE galaxies are selected from the Extended Source Catalog \citep{jarrettetal2000} of the Two Micron All Sky Survey (2MASS) \citep{skrutskieetal2006}. 
The MASSIVE survey is volume limited ($D < 108$ Mpc) and complete down to $M_* \approx 10^{11.5} M_\odot$.

For each galaxy, our observations provide $\sim 750$ fiber spectra
from the Mitchell/VIRUS-P IFS \citep{hilletal2008a} on the 2.7 m Harlan J. Smith Telescope at McDonald Observatory. The IFS covers a $107''$ by $107''$ FOV with 246 fibers, each of $4.1''$ diameter. The fibers are evenly spaced with a one-third filling factor.
We observe three dithering positions to achieve contiguous spatial coverage of each galaxy. The wavelength coverage is 3650\AA\ to 5850\AA\ and the average spectral resolution is 5\AA\ FWHM.  Details of the observing strategy and data reduction procedure are given in \citet{Maetal2014} and \citet{Vealeetal2017a}.

\subsection{Spatial binning and velocity maps}
\label{sub:binning}

We apply the binning procedure described in \citet{Vealeetal2017a}
to ensure that our IFS spectra in the outskirts of galaxies achieve a threshold signal-to-noise ratio (S/N).  The spectra from single fibers in the central regions of our target galaxies always have a S/N higher than the threshold, so no binning is needed here. In the fainter outer parts of each galaxy, we group the fibers with lower S/N into bins until the co-added spectrum reaches the threshold. See \citet{Vealeetal2017a} for details.

To make full use of the velocity map across the entire FOV of each galaxy, we make one modification to the binning scheme of \citet{Vealeetal2017a}.
\citet{Vealeetal2017a} imposed a minimum S/N of 20 for each bin, and to enhance the spatial resolution in the outer parts of each galaxy, the individual fiber spectra with S/N below 20 were first co-added in pairs (i.e. ``folded") over the photometric major axis.
As a result, the velocity maps only contained independent bins for half of the FOV.  Here, we instead generate ``unfolded" maps
and use a lower S/N threshold that is adequate for measuring the line-of-sight velocities.
For ease of comparison with the results in \citet{Vealeetal2017a,Vealeetal2017b,Vealeetal2018},
we opt to preserve the approximate area of each velocity bin in the folded version of the bin maps and therefore impose a minimum S/N of 14 for the unfolded data. This procedure results in roughly twice as many velocity bins as in \citet{Vealeetal2017a}.

To extract the line-of-sight velocity distribution (LOSVD), we use the penalized pixel-fitting (pPXF) method of \citet{cappellariemsellem2004}. This method recovers the LOSVD function $f(v)$ by convolving the observed galaxy spectrum with a set of template star spectra and assuming that $f(v)$ can be reasonably described by a Gauss-Hermite series of order $n=6$:
\begin{equation}
f(v) = \frac{e^{-\frac{y^2}{2}}}{\sqrt{2 \pi \sigma^2}} \bigg[1 + \sum_{m=3}^n h_m H_m(y) \bigg],
\end{equation}
\noindent where $y=(v-V)/\sigma$, $V$ is the mean velocity, $\sigma$ is the velocity dispersion, and $H_m$ is the $m^{\text{th}}$ Hermite polynomial. For details on the parameters of this fitting process, see \citet{Vealeetal2017a}.
For each galaxy, the final result is a set of best-fitting kinematic moments ($V$, $\sigma$ and $h_3$ through $h_6$) for each bin of our unfolded maps.
The error bars on the kinematic moments are estimated using Monte Carlo methods. For each bin, we generate 100 trial spectra where we add random Gaussian noise to the original spectrum -- the Gaussian noise scale for each trial is derived using the actual noise of the spectrum. The reported error represents the standard deviation of the pPXF results for the 100 trial spectra.

To quantify the amount of rotation for each galaxy, we compute the dimensionless spin parameter $\lambda$ \citep{binney2005, emsellemetal2007} at the effective radius $R_e$
(see \citealt{Vealeetal2017a}).  Since all the fibers in the inner regions of the galaxies are treated in the same way in the folded and unfolded versions of our velocity maps, we find $\lambda_e$ measured from the two velocity maps to agree well, differing
by less than $20\%$ for $\sim 90$\% of the galaxies (see columns 3 and 4 of Table~\ref{tab:main}).
Larger differences are found for the small subset of galaxies that exhibit rotation around the photometric {\it major} axis,
which are discussed in Section~\ref{sub:prolate_galaxies}.
This is because the folding procedure in \citet{Vealeetal2017a}
is performed about the photometric major axis under the standard assumption that galaxies rotate around the minor axis.
However, all these galaxies have folded $\lambda_e \lesssim 0.1$, such that the unfolded $\lambda_e$ is still less than 0.1 for most cases (and less than 0.15 for all cases); their classification as slow rotators therefore remains the same.

We observed NGC~4472 early on using a different pointing strategy from that used in the MASSIVE run, and the data are not suitable for the current unfolded analysis pipeline. For completeness, we instead use the parameters derived by \citet{Vealeetal2017a,Vealeetal2017b,Vealeetal2018} and the kinematic results from ATLAS$^\text{3D}$ \citep{krajnovicetal2011}.

\subsection{Photometric data}
\label{sub:wfc3photo}

The photometric position angle (PA$_\text{phot}$) and ellipticity ($\varepsilon$) of each galaxy are taken primarily from the NASA-Sloan Atlas (NSA; http://www.nsatlas.org). For galaxies that do not have NSA photometric data, the 2MASS catalog is used. Since neither NSA nor 2MASS provide error bars on the photometric PA, for the purposes of our analysis we assume a fiducial uncertainty on PA$_\text{phot}$ of 5$\degr$ for all galaxies in our sample.

For a subsample of 35 MASSIVE galaxies, we have obtained detailed WFC3 photometry \citep{Goullaudetal2018} which provides surface brightness profiles and photometric parameters, including luminosity-weighted photometric PA. For 28 of these galaxies, the WFC3 photometric PA agrees with the NSA/2MASS value within $5\degr$ or less. The 7 galaxies that have different WFC3 PAs are NGC~507, NGC~665, NGC~1129, NGC~1272, NGC~1453, NGC~2258, NGC~2274, and NGC~2672. NGC~1272 is a non-rotator (see Section~\ref{sub:non-rotators} for the definition), so the disagreement between the WFC3 and NSA/2MASS photometric position angles does not affect any of our results. For the remaining 6 galaxies, we find that using the WFC3 photometric PA increases the value of the misalignment angle by $\sim 30\degr, 13\degr, 15\degr, 15\degr, 10\degr,$ and $10\degr$, respectively, from the values reported in Table~\ref{tab:main}. However, this does not affect our classification of each galaxy into aligned or misaligned (Section~\ref{sub:aligned-misaligned}) or significantly change our conclusions.


\section{Kinematic Analysis}
\label{sec:kinematic_analysis}

\subsection{Global kinematic position angle PA$_\text{kin}$ and misalignment angle $\Psi$}
\label{sub:pakin_psi}

We use the \texttt{fit\_kinematic\_pa}\footnote{http://www-astro.physics.ox.ac.uk/{\raise.17ex\hbox{$\scriptstyle\sim$}}mxc/software/} routine described in Appendix C of \citet{Krajnovicetal2006} to measure a global kinematic position angle, PA$_\text{kin}$, from the velocity map of each galaxy.
The angle PA$_\text{kin}$ represents the average orientation of the stellar motion in our IFS field of view, and measures the direction of the receding part of the velocity map relative to the north, following the same convention as PA$_\text{phot}$ that eastward is positive.
The routine uses the observed velocity map to generate a bi-anti-symmetric model map for each possible PA$_\text{kin}$ value. It then searches for the value of PA$_\text{kin}$ that corresponds to the best matching model map, defined as the angle that minimizes the $\chi^2$ between the observed map and the model map.

From the photometric position angle PA$_\text{phot}$ and the kinematic position angle PA$_\text{kin}$, we compute the kinematic misalignment angle $\Psi$ following the definition of \citet{Franxetal1991}:
\begin{equation}
  \label{eqn:psi}
	\sin \Psi = |\sin(\text{PA}_\text{phot} - \text{PA}_\text{kin})|.
\end{equation}
We note that while the range of PA$_\text{phot}$ is $0\degr$ to $180\degr$ and the range of PA$_\text{kin}$ is $0\degr$ to $360\degr$, the misalignment angle $\Psi$, by construction, is restricted to be between 0$\degr$ and 90$\degr$.
That is, $\Psi$ measures the misalignment of the two position angles regardless of the sense of the galaxy rotation.

Our measurements of PA$_\text{kin}$ and $\Psi$, along with 1-sigma confidence level errors on PA$_{\text{kin}}$, are listed in Table~\ref{tab:main}.

\subsection{Local kinematic features}
\label{sub:local_kinematic_features}

We use the \texttt{kinemetry}\footnote{http://davor.krajnovic.org/idl/} method \citep{Krajnovicetal2006} to analyse the various local kinematic features in each galaxy in our sample. Kinemetry generalizes isophotal analysis by modeling a galaxy velocity (or higher-order velocity moments) map as simple forms along ellipses: a cosine term for antisymmetric (odd) moments and a constant for symmetric (even) moments. 
The velocity profile along an ellipse is Fourier decomposed as:
\begin{equation}
\label{eqn:kinemetry}
V(a,\psi) = A_0(a) + \sum_{n=1}^N k_n(a) \cos[n(\psi - \phi_n(a))],
\end{equation}
where $a$ is the length of the semi-major axis of the ellipse, $\psi$ is the eccentric anomaly, and $k_n$ and $\phi_n$ are the amplitude and phase coefficients, respectively.

In addition to the kinematic coefficients $k_n$, the kinemetry code returns a local position angle $\Gamma$ and a parameter $q_{\rm kin}$ for the flattening of the best-fitting ellipses along which velocity extraction was performed -- their radial profiles can be used to identify kinematic features on the velocity maps. Briefly, the kinemetry algorithm consists of two steps. At each radius $a$, first, a kinematic profile is extracted for each value of ($\Gamma$, $q_{\rm kin}$) chosen from a finely-sampled grid. The best-fitting  $\Gamma$ and $q_{\rm kin}$ are the ones that minimize $\chi^2 = k_1^2 + k_2^2 + k_3^2$. Then, in the second step, a new Fourier decomposition is performed along the ellipse given by the best-fitting $\Gamma$ and $q_{\rm kin}$ of the previous step.

The physical meanings of the kinemetry parameters are as follows:
\begin{itemize}
	\item The local kinematic position angle $\Gamma$ gives the orientation of the velocity map at each radius. Similar to the global PA$_{\text{kin}}$, it is measured east of north to the receding part of the velocity map. The value $\Gamma = 0$ corresponds to the receding part being aligned with the north direction. As noted in \citet{Krajnovicetal2006}, the global PA$_\text{kin}$ described in Section~\ref{sub:pakin_psi} can be used as a proxy for a luminosity-weighted average $\Gamma$.
We have verified that our global PA$_\text{kin}$ is indeed consistent with the spatial average of our local $\Gamma$.
	\item The flattening $q_{\rm kin}$ (also the ratio of the semi-minor to semi-major axes of the ellipse) is related to the opening angle of the iso-velocity contours. Setting $q_{\rm kin}=1$ translates to velocity extraction along best-fitting circles.
	\item $A_0$ is related to the systemic velocity of the galaxy.
	\item $k_1$ represents the amplitude of the rotational motion.
	\item $k_5$ represents higher-order deviations from the assumption of a simple
cosine law rotation. A high $k_5$ term generally indicates the existence of multiple kinematic components. In practice, one works with the dimensionless ratio $k_5/k_1$.
\end{itemize}

For the galaxies in our samples, we first let kinemetry perform velocity extraction along best-fitting ellipses. This is well suited for the velocity profiles of fast rotators. Applying kinemetry to slow rotators, however, is inherently more complicated since their low rotation velocities can introduce degeneracies in the position angle and flattening parameters. In order to reduce the degeneracies, we rerun the analysis for slow rotators and restrict kinemetry to perform velocity extraction along best-fitting circles. For some intermediate cases, we run kinemetry along ellipses ``restricted'' such that $q_{\text{kin}} \geq 1- \varepsilon_{\text{phot}}$.
We find this to be helpful, for example, for cases where either the velocity is small in the central region and large in the outer region, or vice versa. In both cases, an ellipse expansion across the entire radial range does not describe well the changes in the observed velocity map.


\section{Global Kinematic Alignment}
\label{sec:global_kinematic_alignment}

\subsection{Non-rotators}
\label{sub:non-rotators}

For 19 galaxies (21\%) in our sample, the kinemetry code is unable to identify a well-defined PA$_\text{kin}$, i.e., the error on PA$_\text{kin}$ is close to $90\degr$. We have visually inspected the velocity map of each galaxy and verified that none shows clear signs of organized rotation. All 19 galaxies have low spin with $\lambda_e \la 0.08$ and low rotation velocity with $k_1 \la 30$ km s$^{-1}$, and all were classified as slow rotators in \citet{Vealeetal2017a,Vealeetal2017b,Vealeetal2018}. Here we refer to this subset of slow rotators as {\it non-rotators} (labelled as "None" in column 9 of Table~\ref{tab:main}), and do not include them in the alignment versus misalignment statistics below.

It is possible that the lack of apparent rotation is a projection effect caused by the angular momentum pointing along the line-of-sight. For a population of galaxies with randomly distributed intrinsic angular momentum direction, we expect only $\sim 5\%$ to have their angular momentum pointing along or close to the line-of-sight. 
If this projection effect were indeed the only cause of observed non-rotation,
we would expect to find a non-rotator fraction much smaller than the observed 21\%.  In addition, if the angular momentum is indeed pointing along the line-of-sight (i.e. we are seeing the galaxies face-on), we would expect the non-rotators to have rounder $\varepsilon$. However, only 4 of the 19 non-rotators have ellipticity below 0.1, and 9 have ellipticity between 0.26 and 0.44 (see middle panel of Fig.~\ref{fig:psi_mk_eps_lam}), making it unlikely that all the non-rotators are face-on.  We therefore conclude that most of the non-rotators are indeed rotating very slowly.

\subsection{Rotators: aligned versus misaligned}
\label{sub:aligned-misaligned}

\begin{figure}
  \includegraphics[width=\columnwidth]{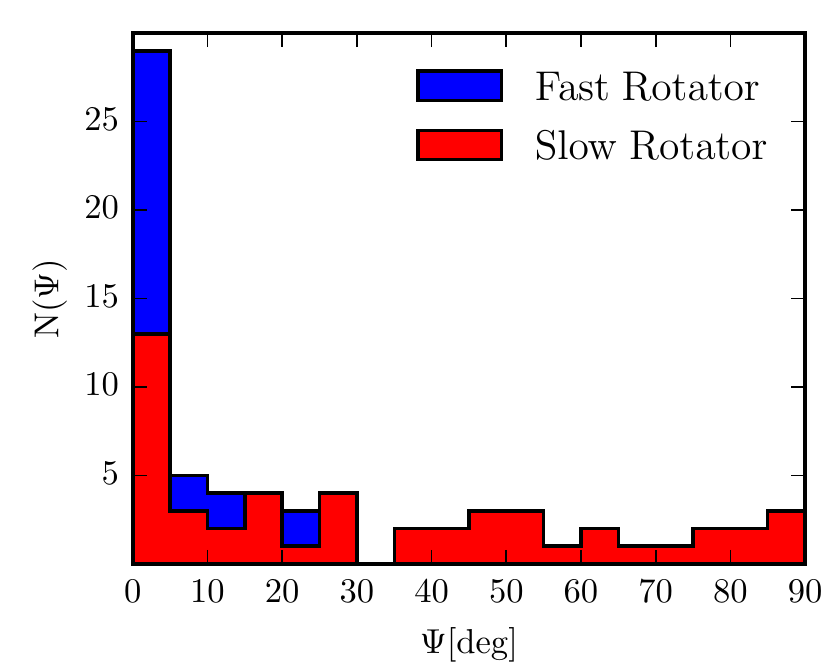}
    \caption{Histogram of the misalignment angle, $\Psi$, between the global kinematic axis PA$_\text{kin}$ and photometric axis PA$_\text{phot}$. The 71 galaxies in our sample of 90 galaxies with measurable rotation and PA$_\text{kin}$ are shown.
We find 91\% (20 out of 22) of the fast rotators (blue) are well aligned with $\Psi < 15\degr$, while only 43\% (21 out of 49) of slow rotators (red) are aligned, with the remaining 57\%
spanning a flat distribution in the full range of misalignment angle. In total, 30 of the 71 galaxies (42\%)
are misaligned with $\Psi > 15 \degr$.}
    \label{fig:kin_misalign}
\end{figure}

In our sample of 90 galaxies, 71 have identifiable kinematic axes. Our measurements of the global kinematic position angle, PA$_\text{kin}$, and the misalignment angle between the photometric and kinematic PA, $\Psi$, for each galaxy are given in Table~\ref{tab:main}.  The distribution of $\Psi$ (Fig.~\ref{fig:kin_misalign}) is sharply peaked at small $\Psi$, with a long and flat tail to maximal misalignment of $90\degr$.  The $\Psi$ distributions for fast and slow rotators are markedly different from each other,
which will be discussed further in Section~\ref{sub:kinematic_misalignment_versus_galaxy_spin}.

Based on the $\Psi$ distribution, we divide the 71 galaxies into aligned versus misaligned, using a similar criterion as in \citet{krajnovicetal2011} and \citet{Fogartyetal2015}:

\begin{itemize}
  \item 41 galaxies (46\% of the total sample) are aligned, for which $\Psi < 15\degr$. Fig.~\ref{fig:aligned} shows the photometric (black bar) and kinematic (red-blue bar) axes for each galaxy.  Due to the larger error bars on PA$_\text{kin}$ for some of the galaxies, we also include in this category 3 galaxies for which $\Psi > 15\degr$ but the error bars of PA$_\text{kin}$ and PA$_\text{phot}$ overlap.
  \item 30 galaxies (33\%) are misaligned, for which $\Psi > 15\degr$ and the error bars of PA$_\text{kin}$ and PA$_\text{phot}$ do not overlap. Fig.~\ref{fig:misaligned} shows how the two axes are misaligned for each galaxy in this category.
\end{itemize}

Galaxy merger simulations have found that gas-rich major mergers of disc galaxies preferentially form axisymmetric oblate remnants with small $\Psi$, whereas gas-poor major mergers of disc galaxies form more triaxial or prolate remnants with a broad distribution of $\Psi$ (e.g., \citealt{NaabBurkert2003, Coxetal2006,Jesseitetal2009}).
The general trends of the simulated $\Psi$ distributions shown in, e.g., fig.~12 of \citet{Moodyetal2014}, figs.~5 and 10 of \citet{Jesseitetal2009} and fig.~13 of \citet{Coxetal2006} are in broad agreement with our results in Fig.~\ref{fig:kin_misalign}.  However, these simulated galaxies are from idealized binary mergers and do not represent a well-defined galaxy population of a given mass range in a cosmological setting.  
Furthermore, these simulations mostly focused on mergers of disc galaxies, which are unlikely to result in galaxies as massive as those in the MASSIVE survey ($M_* \ga 10^{11.5} M_\odot$). 
Further mergers of smaller elliptical galaxies and significant mass accretion are needed to build up such high stellar masses (e.g., \citealt{BKetal2005,BKetal2006}).
Larger simulations probing galaxies in the mass range of MASSIVE would be needed for a more quantitative comparison of Fig.~\ref{fig:kin_misalign} and theoretical predictions.

\subsection{Galaxies with large misalignment angles: minor-axis rotation}
\label{sub:prolate_galaxies}

As Fig.~\ref{fig:kin_misalign} shows,
11 galaxies in our sample have $\Psi > 60\degr$.
In particular, 7 of the 11 galaxies have nearly orthogonal photometric and kinematic position angles with $\Psi > 75\degr$: NGC~708, NGC~1060, NGC~2783, NGC~2832, NGC~7265, NGC~7274, and UGC~2783.
While the rotation of a galaxy is typically found to be
around the photometric minor axis,
these galaxies rotate around the major axis,
sometimes referred to as showing
``minor-axis rotation'' or ``prolate rotation''
(e.g., \citealt{SchechterGunn1979, DaviesBirkinshaw1986, DaviesBirkinshaw1988, Wagneretal1988, JedrzejewskiSchechter1989, Franxetal1989}).

Our data detect rotation around the photometric major axis to varying radii.  The radial extent is $15''-20''$ for NGC~708, NGC~2258, NGC~3209 and UGC~2783, $30''-40''$ for NGC~7265 and NGC~7274, and $30''$ for NGC~1060. NGC~2783 shows minor-axis rotation up to $R\sim 25''$ and mild major-axis rotation in the outer parts. NGC~2832 shows only mild minor-axis rotation up to $R\sim20''$.

Another MASSIVE galaxy with known rotation around the major axis is NGC 7626 \citep{DaviesBirkinshaw1988}. Our detailed velocity map for this galaxy shows hints of
rotation around the major axis in the inner part up to $R\sim20''$, but the outer part up to $R\sim45''$ is rotating around the minor axis at higher velocity. The high velocity rotation in the outer parts affects the global value of the misalignment angle such that $\Psi = 8.0\degr$.

\citet{Tsatsietal2017} present evidence for minor-axis rotation for eight massive galaxies from the CALIFA survey \citep{walcheretal2014}. This rotation occurs either in a main galaxy body or in the kinematically decoupled component of a galaxy. They argue that prolate rotation among massive early-type galaxies is more common than previously thought, citing a fraction of $\sim27\%$ for CALIFA and $\sim23\%$ for ATLAS$^\text{3D}$ galaxies with $M_* \ga 10^{11.3} M_\odot$.
At the high mass end, \citet{Krajnovicetal2018} find that 44\% of their sample of 25 M3G galaxies more massive than $10^{12} M_\odot$ display main rotation around the major axis.
This fraction increases to 50\% if they select only the BCGs in the sample.

Higher-mass galaxies in numerical simulations are found to have a higher probability of being intrinsically prolate or showing minor-axis rotation \citep{EbrovaLokas2017,Lietal2018}.
The emergence of prolate shape and minor-axis rotation typically occurs at similar time \citep{EbrovaLokas2017}, and
the angular momentum of the minor-axis rotation often results from the spin of the primary progenitor in the last significant merger \citep{Lietal2018}.

Keeping all this in mind, we find that 11 out of 90 galaxies exhibit rotation around their respective photometric major axis to some extent, corresponding to a fraction of $\sim 12\%$ in the mass range $M_* \gtrsim 10^{11.5} M_\odot$.

\subsection{Kinematic misalignment versus galaxy spin}
\label{sub:kinematic_misalignment_versus_galaxy_spin}

\begin{figure*}
    \includegraphics[width=\textwidth]{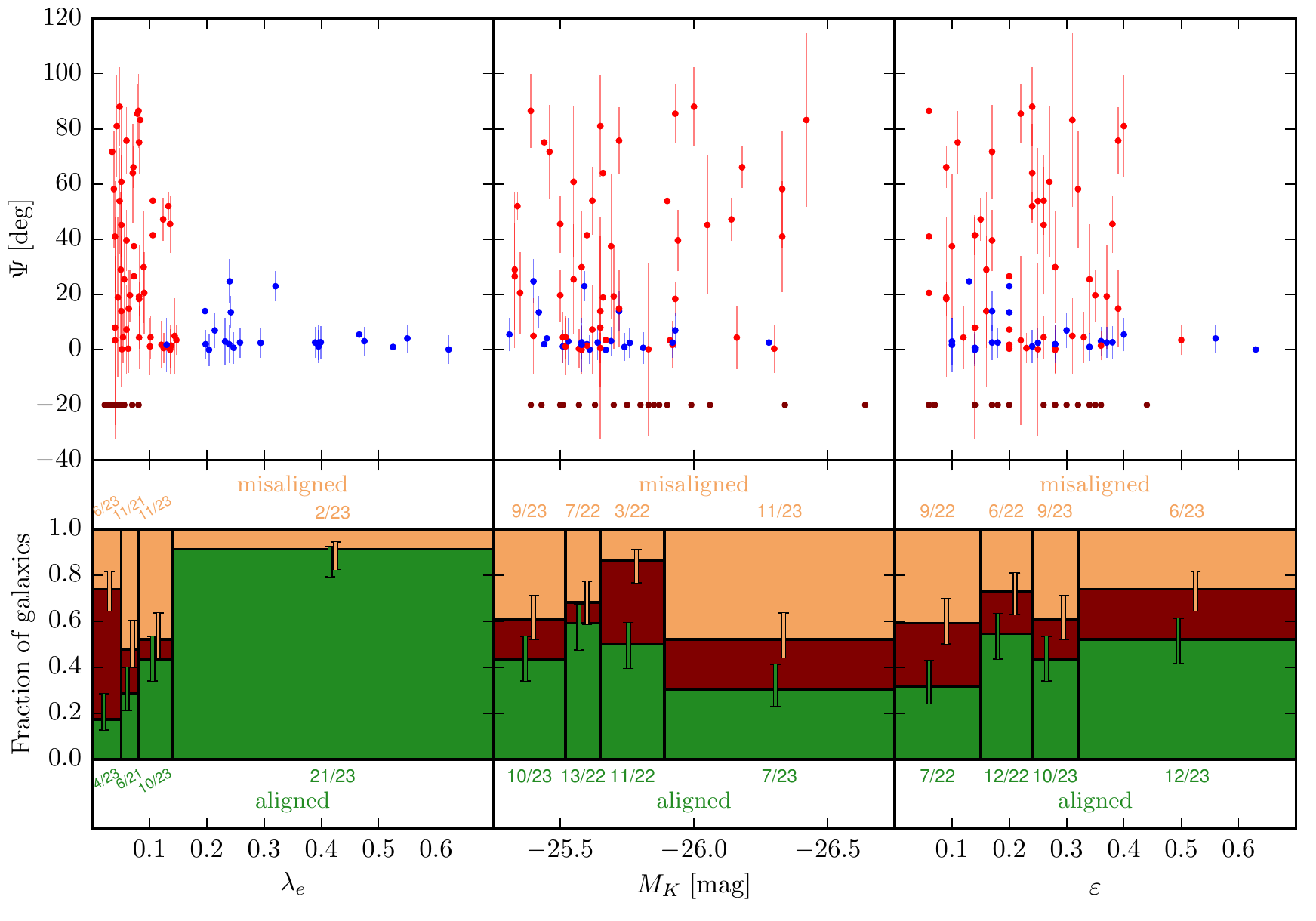}
    \caption{Misalignment angle $\Psi$ versus galaxy properties for the 90 galaxies in our sample. Top: $\Psi$ as a function of spin parameter $\lambda_e$ (left), absolute $K$-band magnitude $M_K$ or stellar mass $M_*$ (middle) and ellipticity $\varepsilon$ (right).  Colors indicate galaxy spin properties: non-rotators (dark red), slow rotators (red), and fast rotators (blue). The non-rotators by definition do not have measurable $\Psi$ and are displaced to the bottom part of each panel. Bottom: Fraction of galaxies that are aligned ($\Psi < 15\degr$; green), misaligned ($\Psi > 15 \degr$; orange), and non-rotating (dark red), in equal galaxy number bins of $\lambda_e$, $M_K$ and $\varepsilon$. The fraction of aligned galaxies decreases noticeably with decreasing spin.
    We do not find any significant correlation with galaxy mass or ellipticity.}
\label{fig:psi_mk_eps_lam}
\end{figure*}

\begin{figure}
    \includegraphics[width=\columnwidth]{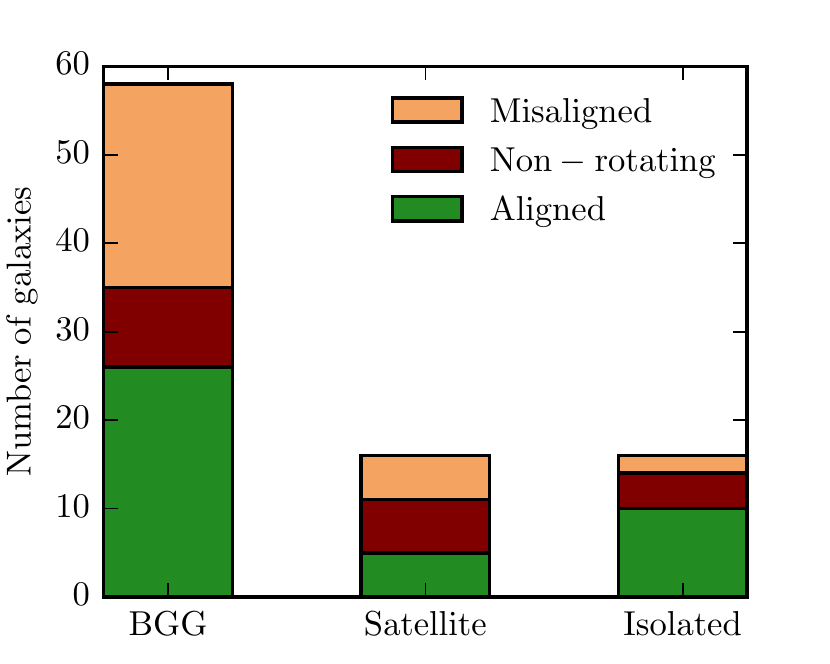}
    \caption{Galaxy alignment versus group membership. For the brightest group galaxies (BGGs) and satellites in our sample, we find a comparable number of aligned (green) vs. misaligned (orange) galaxies. For isolated galaxies, we notice more aligned than misaligned galaxies.}
    \label{fig:align_group}
\end{figure}

\begin{figure*}
    \includegraphics[width=\textwidth]{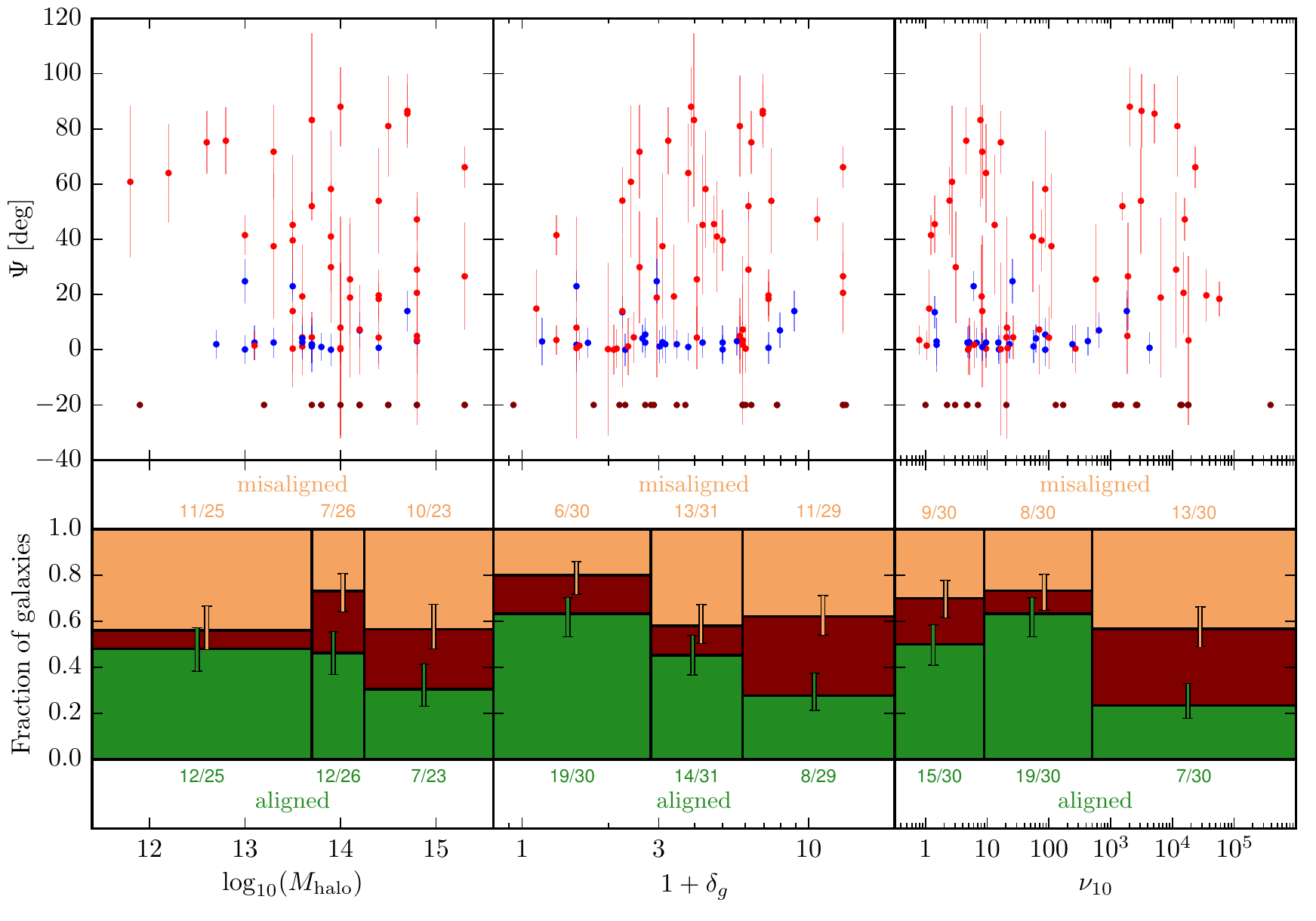}
    \caption{
    Top: Misalignment angle $\Psi$ as a function of galaxy environment measured by halo mass $M_{\rm halo}$ (left), large-scale density $\delta_g$ (middle) and local density $\nu_{10}$ (right).
    Bottom: Fractions of aligned, misaligned and non-rotating galaxies in equal galaxy number bins of $M_{\rm halo}$, $\delta_g$ and $\nu_{10}$.
    The color scheme is the same as in Fig.~\ref{fig:psi_mk_eps_lam}.
    The fraction of aligned galaxies is smaller in denser environments than in less dense ones. There is a noticeable decrease in the aligned fraction with increasing large-scale density.
}
    \label{fig:psi_envir}
\end{figure*}

A total of 22 galaxies in our sample of 90 galaxies have spin parameter $\lambda_e \gtrsim 0.2$, which we classify as fast rotators \citep{Vealeetal2018}. Of the 22 fast rotators, 20 galaxies (91\%) are well-aligned with $\Psi \lesssim 15\degr$ (blue histogram in Fig.~\ref{fig:kin_misalign} and blue points in Fig.~\ref{fig:psi_mk_eps_lam}). The two exceptions are NGC~997 and NGC~6223, with $\Psi \approx 25\degr$ and $23\degr$,
respectively, which we discuss in more detail in Section~\ref{sub:kinematic_twists}.

By contrast, only 21 (43\%) of the 49 slow rotators (excluding the 19 non-rotators) are aligned with $\Psi < 15\degr$. The remaining 28 slow rotators show a wide variation in misalignment angle (red histogram in Fig.~\ref{fig:kin_misalign}), with $\Psi$ distributed fairly evenly from $15\degr$ to $90\degr$.   We therefore conclude that only 21/68 (31\%) of the MASSIVE galaxies with low spins ($\lambda_e < 0.2$) have aligned kinematic and photometric axes.

To verify that the fractions of aligned galaxies among fast and slow rotators are statistically different, we use Fisher's test.
Since there is some uncertainty in assigning a fast/slow rotator label (particularly for galaxies close to the threshold line), as well as an aligned (or not) label, we perform a Monte Carlo (MC) run of Fisher's test where we draw from the uncertainty distributions (in $\lambda_e$ and $\Psi$) for each galaxy	prior to assigning FR/SR and alignment labels.
The resulting distribution of $\log(p-{\rm value})$ is well described by $\mathcal{N}(-5.39, 1.56)$,
confirming that the two fractions are different to at least the 99\% confidence level.
Additionally, we use standard non-parametric tests to verify that the distributions in stellar mass and distance for fast/slow rotators are not statistically different.
This shows that the difference in alignment fraction between fast and slow rotating MASSIVE galaxies is statistically significant and does not result from differences in distance or mass distributions between the two classes.

As can be seen in the left panels of Fig.~\ref{fig:psi_mk_eps_lam}, we find that the misalignment angle $\Psi$ decreases quickly with increasing $\lambda_e$. 
In addition, we find a noticeable increase in the fraction of aligned galaxies with increasing spin parameter, from $17^{+11}_{-4}\%$ in the lowest spin bin to $91^{+2}_{-12}\%$ in the highest spin bin.
An MC run (again, drawing from the uncertainty distributions of $\lambda_e$ and $\Psi$) of a two-sample Kolmogorov-Smirnov (K-S) test 
yield a $p$-value $\ll 0.01$ (with $\log(p-{\rm value}) \sim \mathcal{N}(-4.75, 1.49)$)
and further confirms that there is a significant difference between the spin distribution of the aligned and misaligned samples.

The error bars on the fraction of aligned/misaligned galaxies as a function of $\lambda_e$ (as well as in bins of morphological and environmental parameters) are computed within a Bayesian framework.
The likelihood function for this particular type of problem is a binomial distribution of finding $M$ aligned (or misaligned) galaxies in a sample of size $N$.
For computational simplicity, we assume that the prior is a Beta($\alpha$, $\beta$) distribution of mean $\mu=\alpha/(\alpha+\beta)$ and sample size $n=\alpha+\beta$.
For each sample, we assume that the prior mean is the sample mean (i.e. 41/90 for the aligned sample and 30/90 for the misaligned sample).
We use $n=3$ as the prior sample size for both aligned and misaligned samples.
By Bayes' theorem, the posterior is also a beta distribution with parameters ($\alpha'$, $\beta'$), where $\alpha' = \alpha + M$ and $\beta' = \beta + M - N$.
Finally, we determine the error bars by finding the region that contains 68.3\% of the posterior probability distribution.

In previous kinematic studies of lower-mass galaxies, \citet{Fogartyetal2015} found that for 80 early-type galaxies in the SAMI survey, 83\% of the 62 FRs and 38\% of the 16 SRs are aligned.
For the 260 ATLAS$^\text{3D}$ early-type galaxies, \citet{krajnovicetal2011} found 96\% of the 224 FRs and 56\% of the 36 SRs to be aligned.
While only 24\% of the galaxies in our sample are FRs due to the higher mass range ($M_* \ga 10^{11.5} M_\odot$) probed by the MASSIVE survey, we also find fast rotators to be predominantly aligned.
The non-fast rotators at this high mass, however, show a diverse misalignment behavior.

In a similar spirit to \citet{Franxetal1991}, \citet{Weijmansetal2014} and \citet{Fosteretal2017} inferred the distributions of the intrinsic galaxy shapes from the distributions of the observed ellipticity and kinematic misalignment in the ATLAS$^{\rm 3D}$ and SAMI surveys.
The general finding is that galaxies with high spin are predominantly axisymmetric systems formed by gas-rich mergers, while those with low spins are likely to be mildly triaxial formed by gas-poor mergers.
Considering the prevalence of slow rotators and misalignment in the MASSIVE sample, we expect gas-poor minor and major mergers to be the dominant channel for the late-time mass assembly of early-type galaxies in our mass range ($M_* \ga 10^{11.5} M_\odot$).
To verify this, in Section~\ref{sec:intrinsic_shapes} we perform shape inversion to determine the distribution of intrinsic shapes of MASSIVE galaxies.

\subsection{Kinematic misalignment versus galaxy mass and ellipticity}
\label{sub:kinematic_misalignment_versus_galaxy_mass_and_ellipticity}

Our measurements of the kinematic misalignment angle $\Psi$ as a function of the absolute $K$-band magnitude $M_K$ or stellar mass $M_*$ (middle panel) and ellipticity $\varepsilon$ (right panel) are shown in the top row of Fig.~\ref{fig:psi_mk_eps_lam}.
The corresponding fraction of aligned ($\Psi < 15\degr$; green) and misaligned ($\Psi > 15\degr$; orange) galaxies in bins (with roughly equal number of galaxies) of $M_K$ and $\varepsilon$ are shown in the bottom row.
For completeness, the 19 non-rotators without measurable kinematic axis and $\Psi$ are shown in dark red.

The fraction of misaligned galaxies is not significantly higher for the most massive galaxies than for the less massive ones (lower middle panel of Fig.~\ref{fig:psi_mk_eps_lam}).
Indeed, a two-sample KS test for the $M_*$ distribution yields a $p$-value of $0.25$, showing that misalignment likely does not depend on stellar mass. 

All but 6 galaxies in our sample have $\varepsilon < 0.4$.
At a given $\varepsilon$ (below 0.4), the distribution of $\Psi$ covers the entire range of $0\degr$ to $90\degr$.
This trend was also found by \citet{Franxetal1991}.
As a function of $\varepsilon$, we do not find any significant correlations for the fraction of aligned/misaligned galaxies (right panel of Fig.~\ref{fig:psi_mk_eps_lam}). 
Additionally, we do not find any significant difference between the two samples: a K-S test yields a $p$-value of $\sim 0.64$ when comparing the aligned sample to the misaligned one.
At fixed $\varepsilon$, a galaxy has a roughly equal chance of being aligned or misaligned.
This suggests the probability that a galaxy is aligned or misaligned does not depend on the galaxy's ellipticity.

\subsection{Kinematic misalignment versus galaxy environment}
\label{sub:kinematic_misalignment_vs_galaxy_environment}

We now investigate whether there exist any correlations between the amount of misalignment and the environmental properties described in \citet{Vealeetal2017b}: group membership, halo mass $M_\text{halo}$, large-scale density $\delta_g$ (the density field that surrounds the galaxy on $\sim 5$ Mpc scale), and local galaxy density $\nu_{10}$ (the luminosity density of galaxies in a sphere that contains the 10$^\text{th}$ nearest neighbor).

The results for group membership are shown in Fig.~\ref{fig:align_group}.
We find that among the isolated galaxies, there are more aligned {\bf ($10/16 \approx 62^{+9}_{-14}\%$)} than misaligned {\bf ($2/16 \approx 12^{+11}_{-5}\%$)} galaxies.
Among BGGs and satellites, the numbers of aligned and misaligned galaxies are comparable ($26/58 \approx 45^{+6}_{-6}\%$ aligned versus $23/58 \approx 40^{+6}_{-7}\%$ misaligned galaxies among BGGs and $5/14 \approx 36^{+13}_{-10}\%$ aligned versus $5/14 \approx 36^{+11}_{-12}\%$ misaligned galaxies for satellites.)
Previous studies of lower-mass galaxies have shown that environmental interactions lead to misalignment between the kinematic and photometric axes \citep{B-Betal2014,B-Betal2015}.
They analysed isolated and interacting galaxies in the CALIFA survey with $M_* < 10^{11.4} M_\odot$ and found that isolated galaxies are much more likely to be aligned while interacting galaxies are roughly half aligned and half misaligned.

The misalignment results as a function of $M_\text{halo}$ (left), $\delta_g$ (middle), and $\nu_{10}$ (right) are shown in Fig.~\ref{fig:psi_envir}.
We find that the fraction of aligned galaxies is smaller in more dense environments, as quantified by both $\delta_g$ and $\nu_{10}$.
The strongest such trend is with the large-scale density $\delta_g$ (middle panel), where the fraction of aligned galaxies drops from $67^{+3}_{-14}\%$ to $28^{+9}_{-7}\%$ as $\delta_g$ increases.
We do not find any significant trends between misalignment and halo mass.

Within the volume and mass range sampled by the ATLAS$^{\rm 3D}$ survey, no strong correlations between misalignment and environment were found \citep{krajnovicetal2011}.
They did note that misaligned galaxies were predominantly found in environments with intermediate number densities.
We have previously found that the primary driver of kinematic changes in local early-type galaxies is stellar mass rather than environment \citep[and references therein]{Vealeetal2017b}, it is therefore possible that mass is contributing to the trends between misalignment and environment.

\section{Intrinsic shapes}
\label{sec:intrinsic_shapes}

We follow the method presented in \citet{Franxetal1991}, as implemented by \citet{Fosteretal2017}, to infer the intrinsic shape of slow rotating MASSIVE galaxies.
We assume that our galaxies can be approximated as simple ellipsoids with semi-major axes $a\ge b\ge c$.
This allows us to define the following axis ratios: $p=b/a$ and $q=c/a$, such that $0\le q\le p\le1$.
For oblate systems ($p\sim 1$), $p$ and $q$ may be referred to as the intrinsic disc circularity and intrinsic flattening, respectively.
The technique relies on two observables ($\Psi, \varepsilon$), which can be written in terms of the intrinsic kinematic misalignment ($\Psi_{\rm int}$, the angle between the short axis and the angular momentum vector in 3D), the aforementioned intrinsic axis ratios ($p$ and $q$) and the projection angles ($\varphi, \nu$) as follows:
\begin{equation}\label{eq:PsiEps}
\Psi=\Psi(\Psi_{\rm int}, p, q, \varphi, \nu);\ {\rm and} \
\varepsilon=\varepsilon(p,q,  \varphi, \nu).
\end{equation}
Exact equations can be found in \citet{Fosteretal2017}.
For every galaxy, we have 2 observables and 5 variables.

Assuming that galaxies in the MASSIVE survey are randomly oriented on the sky,
fitting for the whole population removes the dependency on $\varphi$ and $\nu$.
We may solve the equations for a reasonable set of simplifying assumptions on the form of the axis ratio distributions and the distribution in intrinsic kinematic misalignments $\Psi_{\rm int}$.
We assume that all galaxies are drawn from a common distribution in intrinsic axis ratios.
We assume a normal distribution for $q$ with mean $\mu_q$ and standard deviation $\sigma_q$, and a log-normal distribution for $p$ such that $Y=\log(1-p)$, with mean $\mu_Y$ and standard deviation $\sigma_Y$ \citep{Padilla08}.

The final assumption is that the intrinsic kinematic misalignment depends solely on the intrinsic shape of the system.
We assume that $\Psi_{\rm int}$ coincides with the viewing direction that generates a round apparent ellipticity.
This is mathematically equivalent to:
\begin{equation}\label{eq:thetaint}
\tan(\Psi_{\rm int})=\sqrt{\frac{T}{1-T}},
\end{equation}
where
\begin{equation}
T=\frac{1-p^2}{1-q^2}
\end{equation}
is the triaxiality parameter defined in \citet{Franxetal1991}.
This implies $\Psi_{\rm int}=90 \degr$ in prolate systems, while requiring alignment (i.e. $\Psi_{\rm int}=0 \degr$) for oblate systems.
Triaxial systems may take intermediate values of $\Psi_{\rm int}$.
We refer to appendix A of \citet{Weijmansetal2014} for a justification of this assumption.

\begin{figure}
\begin{center}
\includegraphics[width=84mm]{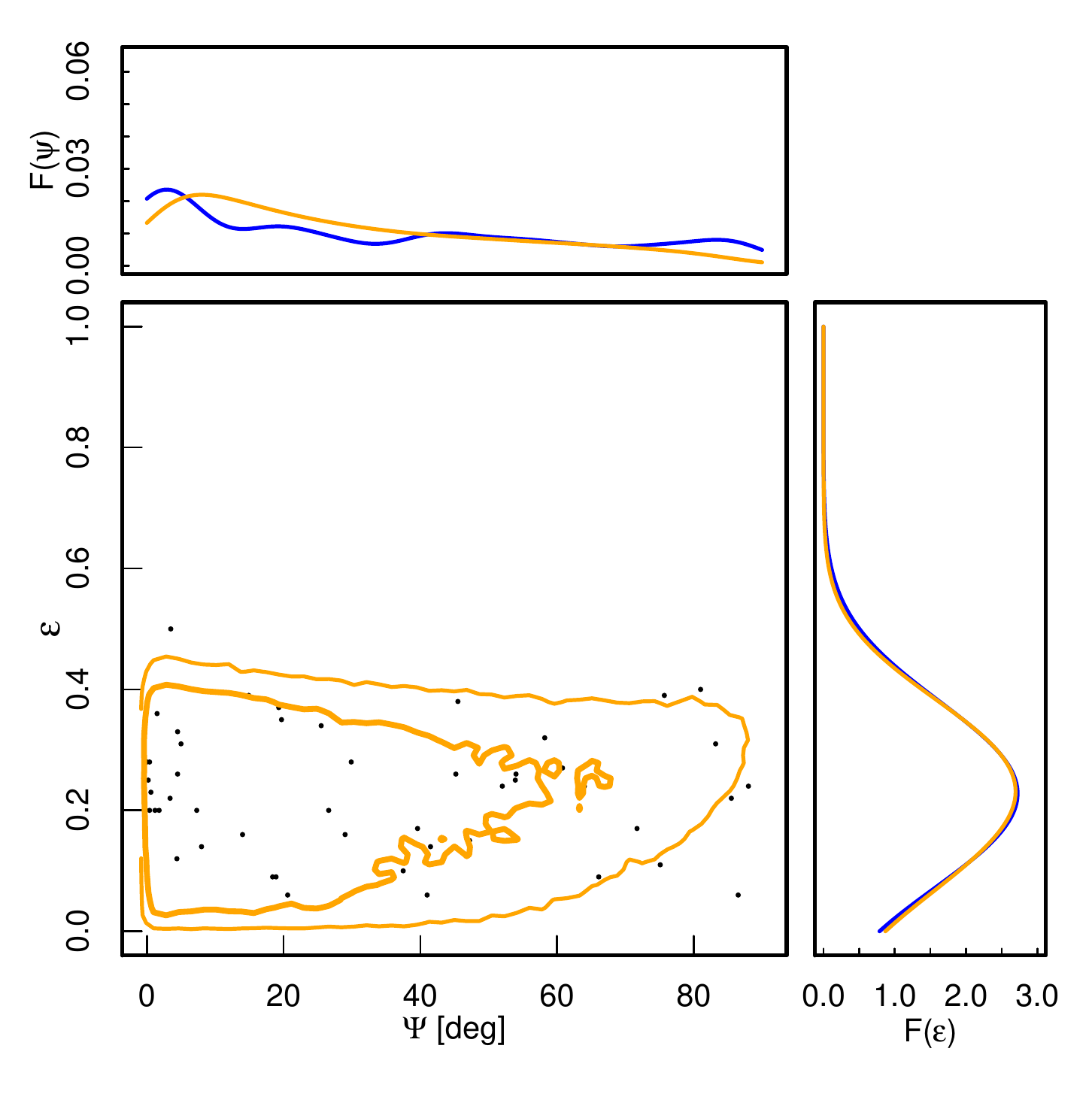}\\ \includegraphics[width=84mm]{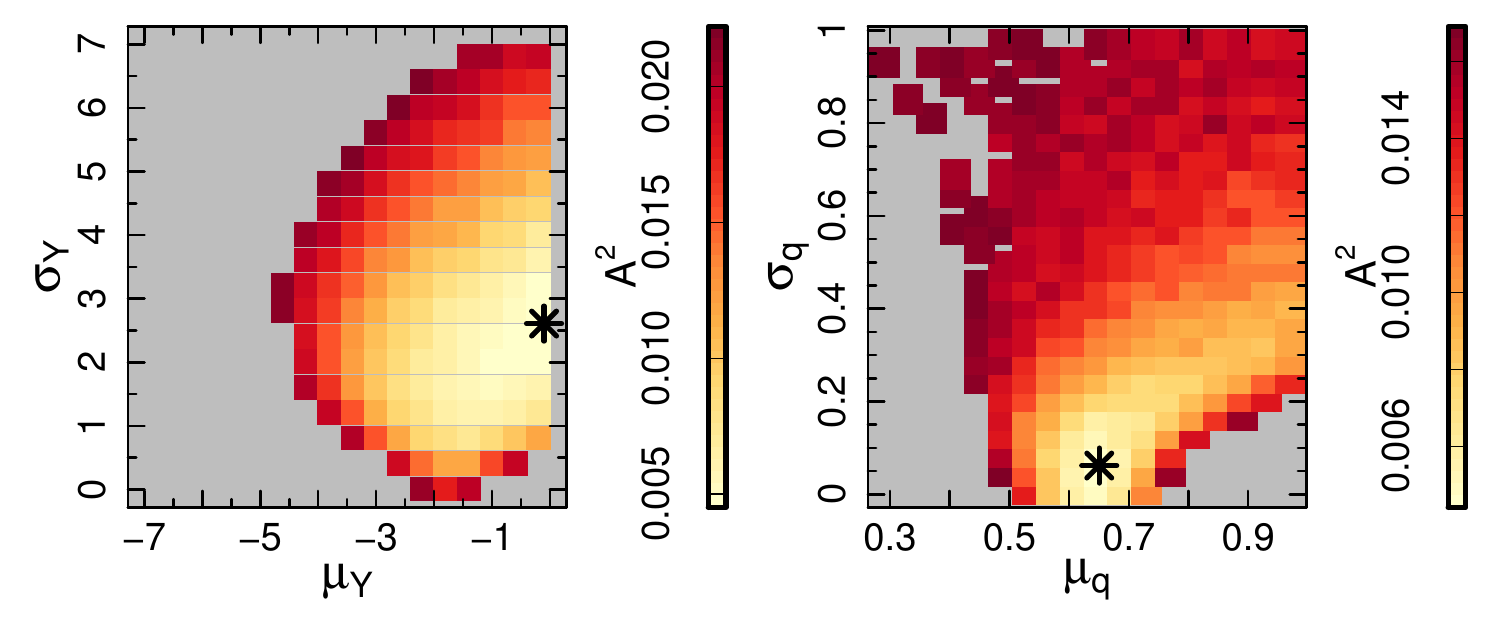}
\caption{The observed distribution of kinematic misalignments $\Psi$ and global apparent ellipticity $\varepsilon$ in MASSIVE slow rotators (top panels). The smoothed and normalised observed (blue) and fitted (orange) distributions $F(\Psi)$ and $F(\varepsilon)$ are shown as separate panels. In the central panels, thick and thin orange lines represent the 68 and 95\%\ probability intervals, respectively, of the 2D distribution. The lower panels show the $A^2$ values in $Y$ vs $\sigma_Y$ space computed by fixing $q$ and $\sigma_q$ to their best fit value or $q$-$\sigma_q$ space (keeping $Y$ and $\sigma_Y$ fixed at their fitted value). In the lower panels, the best fit values are shown as a black asterisk.}\label{fig:shape_MASSIVE}
\end{center}
\end{figure}

\begin{figure}
\begin{center}
\includegraphics[width=84mm]{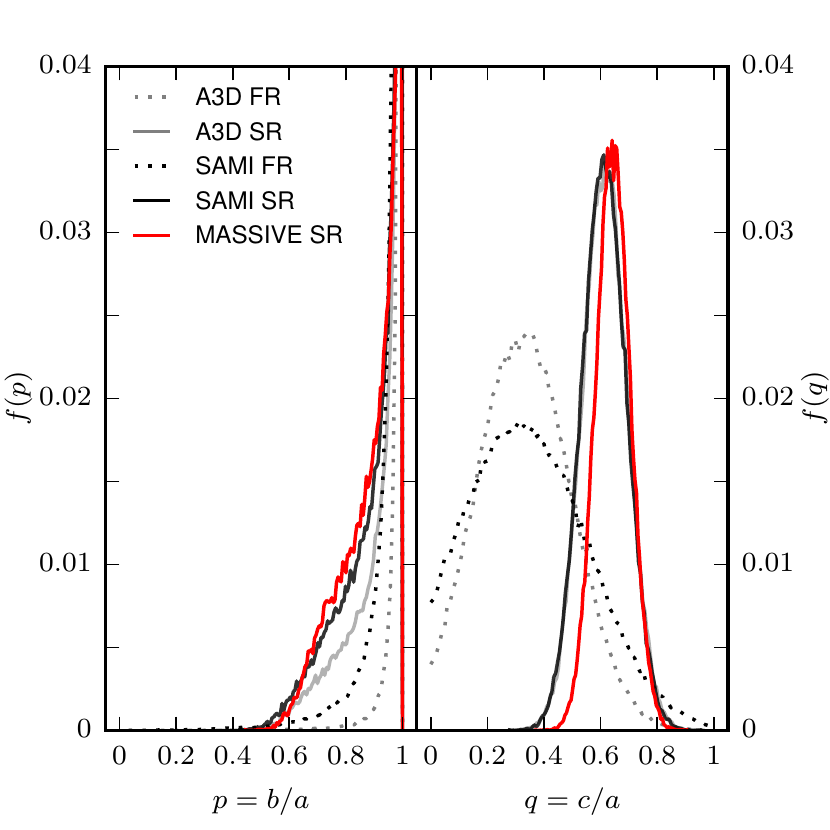}
\caption{
	 Axis ratios distributions for slow (solid lines) rotators in this work (red), and slow and fast (dotted lines) rotators in ATLAS$^{\rm 3D}$ (grey) and SAMI (black). 
	 The lines represent normalised histograms of mock galaxy samples drawn from the best-fitting $p$ and $q$ model distributions. 
Slow rotators have higher proportions of triaxial systems (both $p$ and $q\ne1$), however the typical slow rotator is oblate 
(i.e. the mode of the distribution is $p=1$) 
with a relatively high intrinsic flattening $q\sim0.65$.}\label{fig:lit}
\end{center}
\end{figure}

Fig. \ref{fig:shape_MASSIVE} shows the smoothed normalised distribution of the observables $\Psi$ ($F_{\rm obs}(\Psi)$) and $\varepsilon$ ($F_{\rm obs}(\varepsilon)$) for the MASSIVE slow rotators.
We simultaneously fit the observed distributions by minimising the area between the observed and modelled distributions ($F_{\rm mod}(\Psi)$ and $F_{\rm mod}(\varepsilon)$) using the following parameter:
\begin{equation}\label{eq:chisq} 
\begin{split}
A^2 =  \sum_{i} (F_{\rm obs}(\Psi_{i})-F_{\rm mod}(\Psi_{i}))^2 (\delta \Psi_{i})^{2} + \\ \sum_{j} (F_{\rm obs}(\varepsilon_{j})-F_{\rm mod}(\varepsilon_{j}))^2 (\delta \varepsilon_{j})^{2},
\end{split}
\end{equation}
as per \citet{Fosteretal2017}.
The global minimum $A^2$ is efficiently found using the {\sc DEoptim} R package, which is an implementation of the differential evolution optimisation algorithm \citep[see][for more detail]{Mullen11}.
After 200 iterations, the best fit parameters are $\mu_Y=-0.1$, $\sigma_Y=2.6$, $\mu_q=0.65$, and $\sigma_q=0.06$, corresponding to $A^2=0.0045$.
The intrinsic shape distribution of a mock galaxy sample drawn from the best-fitting log-normal $p$ and normal $q$ model distributions can be visualised in Fig. \ref{fig:lit} along with comparable previous literature studies from the ATLAS$^{\rm 3D}$ \citep{Weijmansetal2014} and SAMI surveys \citep{Fosteretal2017}.
The former was refitted with the same algorithm in \citet{Fosteretal2017}.

From Fig.~\ref{fig:lit} we see that the distribution of the $p$ axis ratio (with mean $p=0.88$ and median $p=0.92$) has a tail towards lower values, indicating that some of the MASSIVE slow rotators are triaxial.
However, 56\% of the galaxies have $p > 0.9$, and these are close to oblate, with an intrinsic flattening of $q \sim 0.65$.

Also visible in Fig.~\ref{fig:lit} is a similar behavior in the inferred intrinsic axis ratios distributions of MASSIVE , ATLAS$^{\rm 3D}$ \citep[e.g., fig. 9 of][]{Weijmansetal2014} and SAMI slow rotators \citep[e.g., fig. 1 of][]{Fosteretal2017}.
Similar to these previous works, we find that the FR/SR difference is larger than differences between SRs from different surveys. 
However, each survey has different stellar mass distributions, so quantitatively determining whether the intrinsic shape distributions depend on $M_*$ within the slow rotator populations will require more careful analysis.


\section{Local Kinematic features}
\label{sec:local_kinematic_features}

The spatial profiles of the local kinematic position angle $\Gamma$ (measured with respect to the global PA$_\text{kin}$) and circular velocity coefficient $k_1$ (defined in Eq.~\ref{eqn:kinemetry}) are two main quantities that can be used to identify local kinematic features in galaxy velocity maps.
For clarification, we note that while our alignment classification in Section~\ref{sec:global_kinematic_alignment} is based on the global PA$_\text{kin}$ and PA$_\text{phot}$, there can be local kinematic variations (e.g., twists, distinct cores) about the global kinematic axis in either aligned or misaligned galaxies.
The goal of this section is to examine these local features.

\subsection{Radial profiles of $\Gamma$ and $k_1$}
\label{sub:radial_profiles}

\begin{figure*}
    \includegraphics[width=\textwidth]{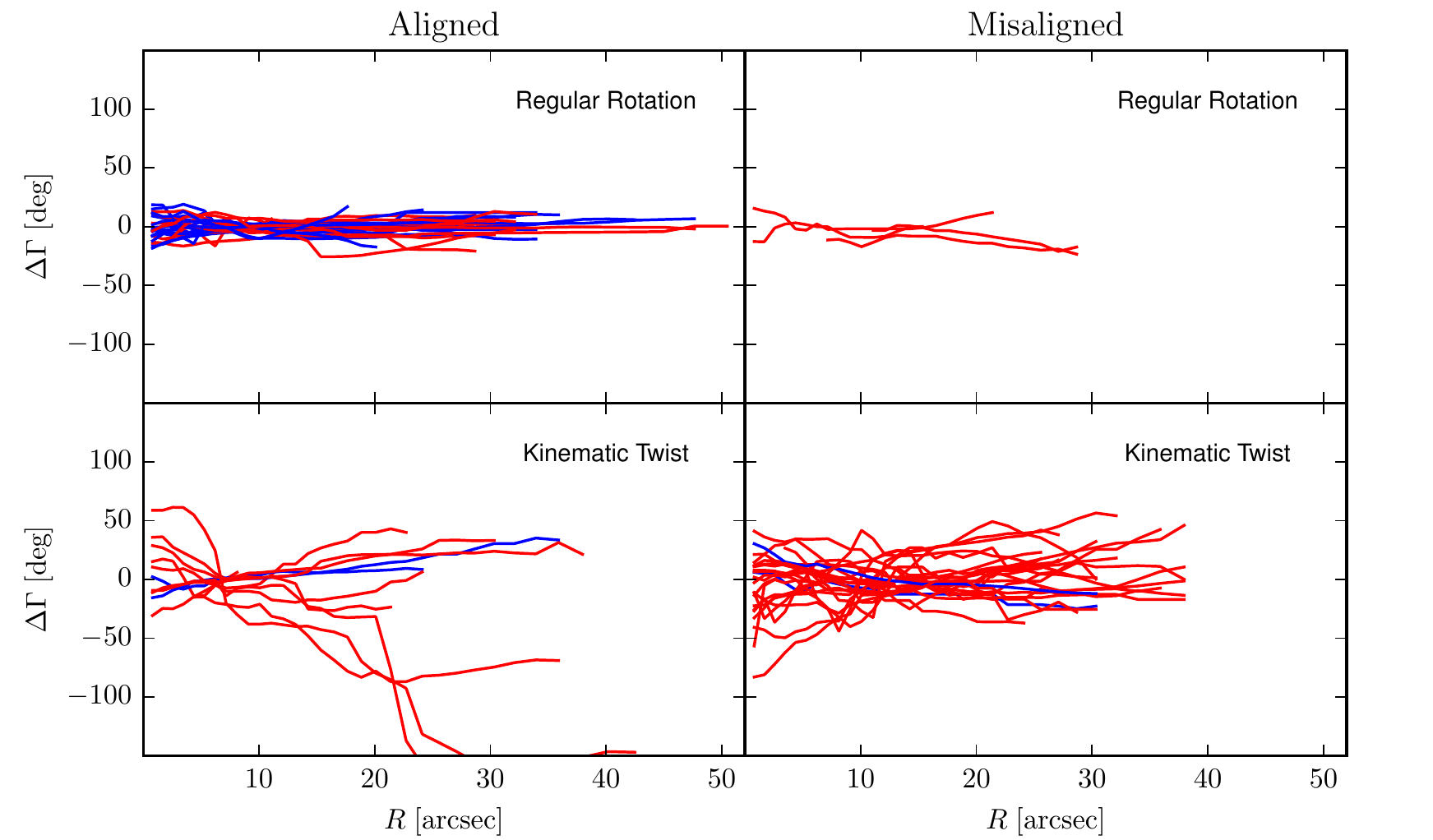}
    \caption{Radial profiles for the local kinematic position angle $\Gamma$ with respect to the global kinematic position angle PA$_\text{kin}$. Regular rotators ($\Gamma$ varying by $< 20\degr$) are mostly (globally) aligned (upper left panel) and kinematic twisters are mostly (globally) misaligned (lower right panel), but 4 misaligned galaxies have nearly constant local kinematic axis (upper right), and 11 aligned galaxies show large local variations in $\Gamma$ (lower left). Fast and slow rotators are color-coded blue and red, respectively.}
    \label{fig:gamma_profiles}
\end{figure*}

Fig.~\ref{fig:gamma_profiles} shows the radial profile $\Gamma(R)$ of the local kinematic position angle for the 41 aligned galaxies (left panels) and 30 misaligned galaxies (right panels) in our sample.
The local kinematic features can be classified into one of the 3 following categories:

\begin{itemize}
  \item Regular rotation: Variations in $\Gamma(R)$ are smooth and less than $\sim 20\degr$ from PA$_\text{kin}$ across the entire map (top panels of Fig.~\ref{fig:gamma_profiles}).
  \item Kinematic twist (KT): Variations in $\Gamma(R)$ are smooth and more than $20\degr$ from PA$_\text{kin}$ (bottom panels of Fig.~\ref{fig:gamma_profiles}; see Section~\ref{sub:kinematic_twists}).
  \item Kinematically distinct components (KDC): $\Gamma(R)$ has an abrupt change of greater than $\sim 30\degr$ over a small region. In addition, $k_1$ decreases and/or $k_5$ increases in the transition region  (Figs.~\ref{fig:0507} and \ref{fig:5322}; see Section~\ref{sub:kinematically_distinct_components}).
\end{itemize}

While the majority of aligned galaxies shows regular rotation (upper left panel of Fig.~\ref{fig:gamma_profiles}) and the majority of misaligned galaxies shows kinematic twists (lower right panel), the correspondence is not one-to-one:  11 of the 41 globally aligned galaxies exhibit local kinematic twists (lower left panel), and 4 of the globally misaligned galaxies have regular rotation about their respective kinematic axis with minimal twists.
Furthermore, we identify one aligned galaxy and one misaligned galaxy exhibiting KDCs. We will discuss more details of kinematic twists and KDCs in the next two subsections.

The outer parts of our velocity maps for 6 galaxies (NGC~0708, NGC~2258, NGC~2832, NGC~3562, NGC~3816, and NGC~6575) show systemic red or blue shifts while the inner regions show clear rotation.
The velocity difference between the outer and inner parts is roughly 10, 20, $-40$, 30, 30, 75 km s$^{-1}$, respectively.
This systemic offset in the galaxy outskirts could be due to low S/N in the outer bins that would affect the pPXF recovered velocity or to a non-existent rotation in the outskirts which kinemetry is ill-suited to analyse.
For these galaxies, we classify the kinematic features based on $\Gamma(R)$ in the inner regions where clear rotation could be detected.

\begin{figure*}
    \includegraphics[width=\textwidth]{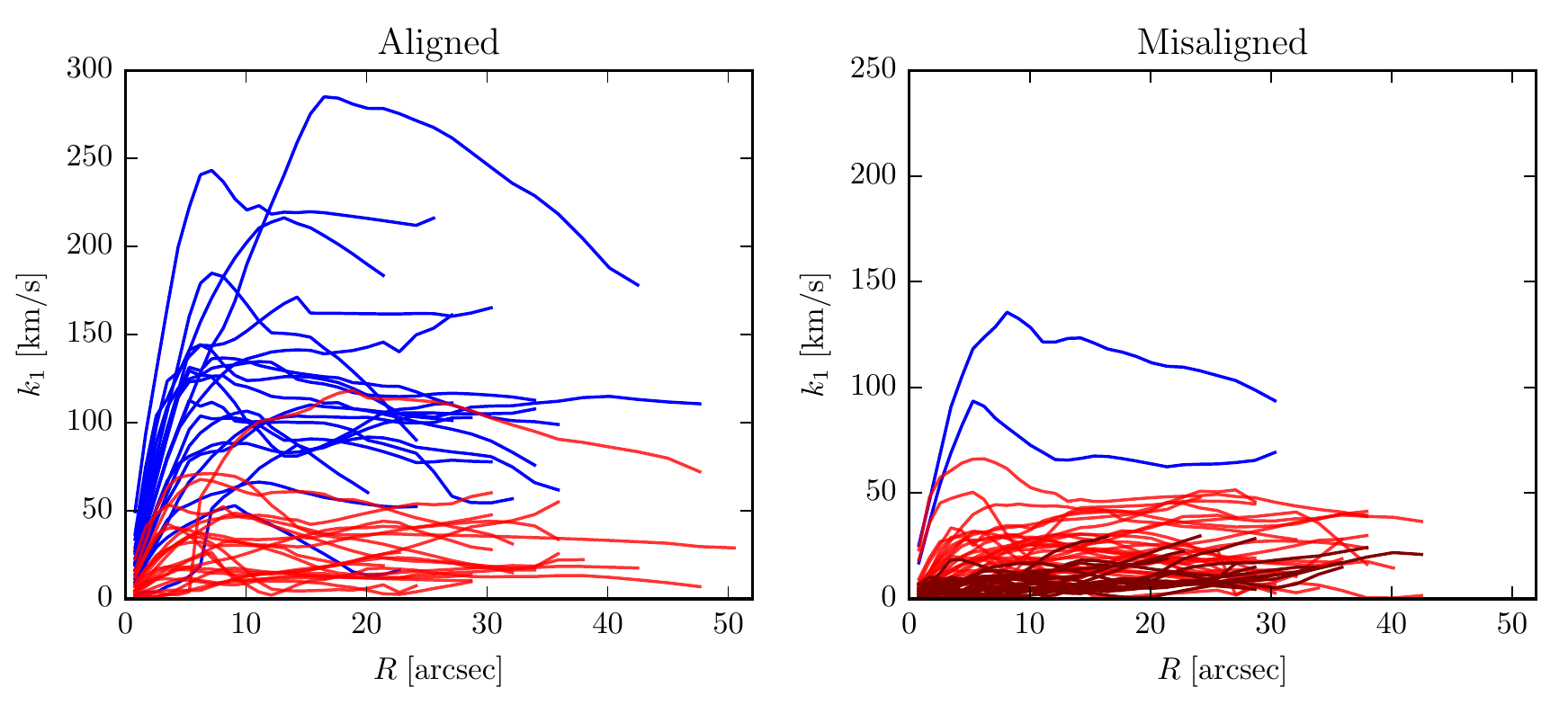}
    \caption{Radial profiles for the $k_1$ velocity coefficient for aligned (left panel) and misaligned (right panel) galaxies. Fast and slow rotators are color-coded in blue and red, as before. The non-rotators are shown in dark red.}
    \label{fig:allSR_k1}
\end{figure*}

The radial profiles of the $k_1$ velocity coefficient are shown in Fig.~\ref{fig:allSR_k1}.
A typical $k_1(R)$ in our sample rises in the inner $\sim 10''$ and then either decreases or stays roughly constant.
With a few exceptions, $k_1 \sim 50$ km s$^{-1}$ is seen to separate fast (blue) from slow (red) rotators (defined by the spin parameter $\lambda_e$), regardless of misalignment.
All non-rotators (dark red) consistently populate the region with $k_1 \lesssim30$ km s$^{-1}$.
The apparent outlier in Fig.~\ref{fig:allSR_k1} (left panel) is the slow rotator (NGC~6482) with $k_1(R)$ rising to $\sim 120$ km s$^{-1}$.
NGC~6482 is nonetheless classified as a slow rotator because its high central $\sigma$ values compensated for the high rotation speeds to lower the spin parameter $\lambda_e$ to 0.14, which is at the high end of the $\lambda_e$ distribution for slow rotators in our sample (see Fig.~6 of \citealt{Vealeetal2017b}).

Table~\ref{tab:summary_kinemetry_features} summarizes the statistics of the kinematic features that we identified in the velocity maps.
For comparison, \citet{krajnovicetal2011} defined 7 types of velocity features for the ATLAS$^{\text{3D}}$ sample, where the data was collected using the SAURON IFS with 0.9$''$ lenses and a $33'' \times 41''$ FOV.
Since we have both a large FOV ($107'' \times 107''$) and larger fibers ($4.1''$ diameter) than ATLAS$^{\text{3D}}$, we can only identify kinematic features that occur on larger scales. For this reason, our kinematic categories are similar, but not identical to the corresponding ATLAS$^{\text{3D}}$ ones.

\begin{table}
  \centering
  \caption{Statistics of local kinematic features \tablefootnote{Excluding NGC~4472 since we do not have detailed kinemetry profiles.}}
  \label{tab:summary_kinemetry_features}
  \begin{tabular}{lccc}
      \toprule
      Feature & Total No. & Aligned & Misaligned \\
      \midrule
      Regular Rotation & 32 &28  &4 \\
      Kinematic Twist & 36 & 11 &25 \\
      Kinematically Distinct & 2 &1 &1 \\
      ~Components & \\
      No Rotation  & 19  & - & -\\
      \bottomrule
  \end{tabular}
\end{table}

\subsection{Kinematic Twists}
\label{sub:kinematic_twists}

We classify galaxies whose $\Gamma(R)$ changes smoothly by more than $20\degr$ from PA$_{\rm kin}$ as displaying a kinematic twist (KT).
We find more than twice as many cases of kinematic twists among the misaligned galaxies (25 misaligned versus 11 aligned), and the twists tend to have larger amplitudes (lower panels of Fig.~\ref{fig:gamma_profiles}).
Out of the 30 misaligned galaxies, only 2 are classified as fast rotators (NGC~997 and NGC~6223), and both galaxies show local kinematic twists (blue curves in the lower right panel of Fig.~\ref{fig:gamma_profiles}).
For NGC~997, $\Gamma$ twists by $\sim 30\degr$, and its PA$_\text{kin}$ ($219.5\degr$) represents the rotation direction of a faster component within the first $\sim 10''$, while the outer parts rotate with a lower velocity and the kinematic axis is aligned with the photometric axis of $14.7\degr$ (within the criterion described in Section~\ref{sec:global_kinematic_alignment}).
NGC~6223 shows a smooth twist of about $40\degr$ over the entire map, while its PA$_\text{kin}$ agrees with the median $\Gamma$.

Among the slow rotators, we find that many slow rotators show large twists of $\sim 30-40 \degr$.
Two examples are NGC~1129 and NGC~4874 for which \citet{Vealeetal2017a} found  misalignment between the photometric and kinematic axes.
We can now quantify the amount of misalignment: NGC~1129 has PA$_\text{kin}=179\degr$ and $\Psi=47\degr$, while NGC~4874 has PA$_\text{kin}=335\degr$ and is misaligned by $66\degr$.
Since the $\Gamma(R)$ profile of NGC~1129 shows a twist of $\sim 30\degr$ across a limited radial range, it would seem that it exhibits a kinematically distinct inner component.
However, because its $k_1$ and $k_5$ profiles do not show the drop and/or rise, respectively, indicative of multiple kinematic components, we classify NGC~1129 as having a kinematic twist.
Meanwhile, NGC~4874 shows a kinematic twist of $\sim 40\degr$ across the entire map, and like NGC~6223, its PA$_\text{kin}$ agrees with the median $\Gamma$, reflecting the misalignment of the galaxy as a whole.
Both galaxies also show significant twists in their photometric axis. \citet{Goullaudetal2018} found that PA$_\text{phot}$ of NGC~1129 has a large twist from $0\degr$ in the inner part (which is very well aligned with the global kinematic axis) to $90\degr$ in the outer part.
The PA$_\text{phot}$ of NGC~4874 twists from $\sim 20\degr$ to $\sim 55\degr$, making it misaligned with the kinematics at all radii.

\citet{Moodyetal2014} analysed the properties of mock kinematic maps of remnants (comparable to ATLAS$^{\rm 3D}$ galaxies) in a set of multiple merger simulations of disc progenitors.
They found that kinematic twists are more common ($\sim 20-80\%$ of projections) in simulated multiple merger remnants  (which resemble observed slow rotators) than in binary mergers which produce fast-rotating remnants ($\sim 10-20\%$ of projections).
For our sample of 90 galaxies, we also find that kinematic twists are more prevalent among slow rotators: $47\%$ of slow rotators have kinematic twists versus $18\%$ of fast rotators (corresponding to a $p$-value of $0.023$ for a Fisher's test).
Since the simulated remnants and our MASSIVE galaxies probe different mass ranges, we do not expect exact quantitative agreement for the occurrence of kinematic twists in fast/slow rotators.
However, the higher prevalence of kinematic twisting among our slow rotators is in agreement with the predictions of simulations.

\subsection{Kinematically Distinct Components in NGC~507 and NGC~5322}
\label{sub:kinematically_distinct_components}

We find kinematically distinct components (KDC) in NGC~507 and NGC~5322.
Both are slow rotators with $k_1^{\rm max} \la 50$ km s$^{-1}$ and $\lambda_e$ below 0.1.
The KDC of NGC~507 was reported in our earlier paper \citep{Vealeetal2017a}.
The KDC of NGC~5322 was reported in \citet{Bender1988} based on long-slit observations and was observed as part of the ATLAS$^{\rm 3D}$ survey \citep{krajnovicetal2011}.
NGC~5322 is one of the six ATLAS$^{\rm 3D}$ galaxies that are massive enough to be in the MASSIVE survey \citep{Maetal2014}.

The velocity map and kinematic parameters for NGC~507 are shown in Fig.~\ref{fig:0507}.
The local kinematic position angle $\Gamma(R)$ shows a large jump of $\sim 100 \degr$ between a radius of $10''$ and $20''$.
The inner kinematic component extends to $R\sim 10''$ and exhibits a disk-like rotation with $k_1\sim 50$ km s$^{-1}$ around a median PA$_\text{kin}$ of $\approx 185\degr$.
The outer component rotates at $\sim 40$ km s$^{-1}$ around a median PA$_\text{kin}$ of $\approx 290\degr$, and it shows a mild twist of $\sim 25\degr$ around this outer axis.
Our measurement of the global PA$_\text{kin}$ for NGC~507 is $183.5\degr\pm 4.0 \degr$, which is close to the the rotation direction of the inner, higher-velocity component.

The global photometric PA for NGC~507 according to the NSA is $21.9\degr$,
but our data for this galaxy from the Hubble Space Telescope (HST) Wide Field Camera 3 (WFC3; F110W filter) reveal a dramatic change of $\sim 70\degr$ in PA$_\text{phot}$ \citep{Goullaudetal2018} that roughly traces the kinematic changes out to $\sim 40"$; see blue versus black points in top left panel of Fig.~\ref{fig:0507}.
Our measured misalignment angle for NGC~507 is $\Psi = 18.4\degr$, which is primarily due to the slight offset of $\sim 15 \degr$ between the photometric and kinematic PA in the inner component.

For the KDC in NGC~5322, Fig.~\ref{fig:5322} shows that the inner kinematic component extends up to radius $\sim 10''$ and is rotating at $k_1 \sim 40$ km s$^{-1}$around a median PA$_\text{kin}$ of $\sim 100 \degr$.
The outer kinematic component is counter-rotating with a lower speed in almost exactly the opposite direction (median PA$_\text{kin}$ of $\sim 270\degr$) as the inner component.
Similar features for the inner $\sim 35''$ were also reported in \citet{krajnovicetal2011}.
The rotation axis for both components is aligned with the photometric axis (PA$_\text{phot}=92\degr$ from NSA) and the misalignment angle is $\Psi=4.5\degr$, so this galaxy is classified as aligned.
Our HST WFC3 data for NGC~5322 shows a constant PA$_\text{phot}$ of identical value as the NSA out to $R\sim 70''$ \citep{Goullaudetal2018}.

We note that the Mitchell IFS used for our MASSIVE survey has both a larger FOV ($107''\times 107''$ vs. $33''\times 41''$) and a larger fiber/lens size ($4.1''$ vs. $0.9''$) than the SAURON IFS used for the ATLAS$^\text{3D}$ survey.
We can therefore only resolve kinematic features above $\sim 5''$, such as the KDCs in NGC~507 and NGC~5322 that occur over a large radial extent of $R \ga 10''$.
Many KDCs in the ATLAS$^\text{3D}$ survey occur on smaller scales and in slow rotators \citep{krajnovicetal2011}.
It is therefore possible that more MASSIVE galaxies have KDCs below the resolution scale of the Mitchell IFS.
On a similar note, NGC~4472 is classified by \citet{krajnovicetal2011} as a KDC with the inner component rotating in the opposite direction of the outer component.
Given the small radial extent of the KDC, it is unlikely that we could have resolved the KDC even if we had the necessary data to produce unfolded maps.
We will report central kinematic features in a subset of MASSIVE galaxies observed with an IFS with $0.2''$ lenslets in a subsequent paper (Ene et al. in preparation).


\section{Conclusions}
\label{sec:conclusions}

In this paper we have studied the stellar velocity features of 90 early-type galaxies in the volume-limited MASSIVE survey, with $M_K \leq -25.3$ mag and $M_* \gtrsim 10^{11.5} M_\odot$.
Our velocity map for each galaxy is constructed from $\sim 750$ closely-packed fiber spectra from the Mitchell IFS (with 3 ditherings) at the McDonald Observatory.
The $107''\times 107''$ FOV of the IFS provides kinematic maps beyond two effective radii for many galaxies in our sample.

We use the kinemetry method of \citet{Krajnovicetal2006} to
identify the average kinematic axis (PA$_\text{kin}$) for each galaxy in our sample (Section~\ref{sec:kinematic_analysis}) and to measure the global misalignment angle $\Psi$ between this axis and the photometric axis (Section~\ref{sec:global_kinematic_alignment}).
We further identify detailed local kinematic features such as twists and distinct cores (Section~\ref{sec:local_kinematic_features}).

Our main findings are:
\begin{itemize}
\item 19 of the 90 galaxies (21\%) are non-rotators that have no detectable rotation or identifiable kinematic axis.
For the other 71 galaxies, 22 are fast rotators (with spin parameter $\lambda_e \gtrsim 0.2$) and 49 are slow rotators.

\item 41 of the 71 rotating galaxies show good alignment between the photometric and kinematic axes ($\Psi < 15\degr$), while 30 galaxies are misaligned ($\Psi > 15\degr$).
Only 2 of the 22 fast rotators (9\%) show misalignment, while by contrast, 28 of the 49 slow rotators (57\%) are misaligned with a nearly flat distribution of $\Psi$ out to $90\degr$ (Fig.~\ref{fig:kin_misalign}).

\item We find 11 galaxies to have large misalignment angle ($\Psi > 60\degr$), 7 of which have $\Psi > 75 \degr$ and are rotating around the photometric major axis rather than the more typical minor axis (Section~\ref{sub:prolate_galaxies}).

\item We find a strong trend of increasing fraction of aligned galaxies with increasing spin parameter, from $17^{+11}_{-4}\%$ in the lowest spin bin to $91^{+2}_{-12}\%$ in the highest spin bin (Section~\ref{sub:kinematic_misalignment_versus_galaxy_spin}  and Fig.~\ref{fig:psi_mk_eps_lam}).
We also find that kinematic misalignment likely does not depend on stellar mass, the fraction of misaligned galaxies at higher mass is not significantly higher than at lower mass.
The fraction of aligned galaxies does not depend on ellipticity (Section~\ref{sub:kinematic_misalignment_versus_galaxy_mass_and_ellipticity}).	

\item The fraction of galaxies with aligned kinematic and photometric axes is smaller in denser galaxy environments, as measured by large-scale galaxy density $\delta_g$ and local galaxy density $\nu_{10}$ (Fig.~\ref{fig:psi_envir}).
The strongest trend is with $\delta_g$, where the fraction of aligned galaxies drops from $67^{+3}_{-14}\%$ to $28^{+9}_{-7}\%$ as $\delta_g$ increases.
As a function of group membership, we find similar numbers of aligned and misaligned galaxies among BGGs and satellite galaxies (Fig.~\ref{fig:align_group}).
But for isolated galaxies, we find 10 aligned ($62^{+9}_{-14}\%$)and only 2 misaligned ($12^{+11}_{-5}\%$) galaxies.
This is consistent with the idea that environmental interactions can be a potential cause of misalignment.

\item We use the kinematic misalignment and ellipticity distributions to infer the intrinsic shape distribution of our sample (Section~\ref{sec:intrinsic_shapes}).
For MASSIVE slow rotators, the mean axis ratios are $p=b/a=0.88$ and $q=c/a=0.65$, consistent with being mildly triaxial (Fig.~\ref{fig:lit}).
The inferred axis ratio distributions have similar trends as those of ATLAS$^\text{3D}$ and SAMI galaxies, however, a more careful analysis is required to determine whether the intrinsic shape distributions depend on stellar mass. 

\item 45\% of the 71 rotating galaxies exhibit regular rotation in which the local kinematic position angle, $\Gamma(R)$, varies by less than $20\degr$ across the entire velocity map, while 51\% show twisting of the kinematic axis with $\Gamma(R)$ varying by more than $20\degr$ (Section~\ref{sec:local_kinematic_features} and Fig.~\ref{fig:gamma_profiles}).
Most of the regular rotators are aligned (on average), and most of the local twisters are misaligned (on average), but the correspondence is not one-to-one (off-diagonal panels in Fig.~\ref{fig:gamma_profiles}).
We identified kinematically distinct components in NGC~507 and NGC~5322 (Section~\ref{sub:kinematically_distinct_components} and Figs.~\ref{fig:0507} and \ref{fig:5322}).

\end{itemize}

We have shown in this paper that the remarkable near alignment of the photometric and kinematic axes found in lower-mass early-type galaxies in the ATLAS$^\text{3D}$ survey does not extend to higher-mass early-type galaxies.
The fast rotators in the MASSIVE survey are indeed very well aligned, consistent with the expectation that they have oblate shapes and rotate around the photometric minor axis.
Unlike ATLAS$^\text{3D}$, however, the majority of the galaxies in our sample are slow rotators, among which 57\% are misaligned with $\Psi$ evenly distributed from $15\degr$ to $90\degr$.  This flat distribution can be produced by
a population of intrinsically (mildly) triaxial galaxies.

In the context of galaxy formation, the high frequency of kinematic misalignment and the broad distribution of the misalignment angle of slow rotators reported in this paper support the notion that major gas-poor mergers are a primary formation process leading to mildly triaxial shapes in massive early-type galaxies.
Our study demonstrates the importance of using volume-limited samples with well-defined galaxy mass range when making inferences about the properties or formation history of early-type galaxies.

\section*{Acknowledgements}

We thank Marijn Franx for enlightening discussions.
The MASSIVE survey is supported in part by NSF AST-1411945, NSF AST-1411642, HST-GO-14210, and HST-AR-14573.
Part of this research was conducted by the Australian Research Council Centre of Excellence for All Sky Astrophysics in 3 Dimensions (ASTRO 3D), through project number CE170100013.




\bibliographystyle{mnras}
\bibliography{massive_x}




\appendix

\section{Kinemetry results and 2D velocity maps for galaxies with kinematically distinct components}

\begin{figure*}
	\vspace*{-0.02in}
	\includegraphics[]{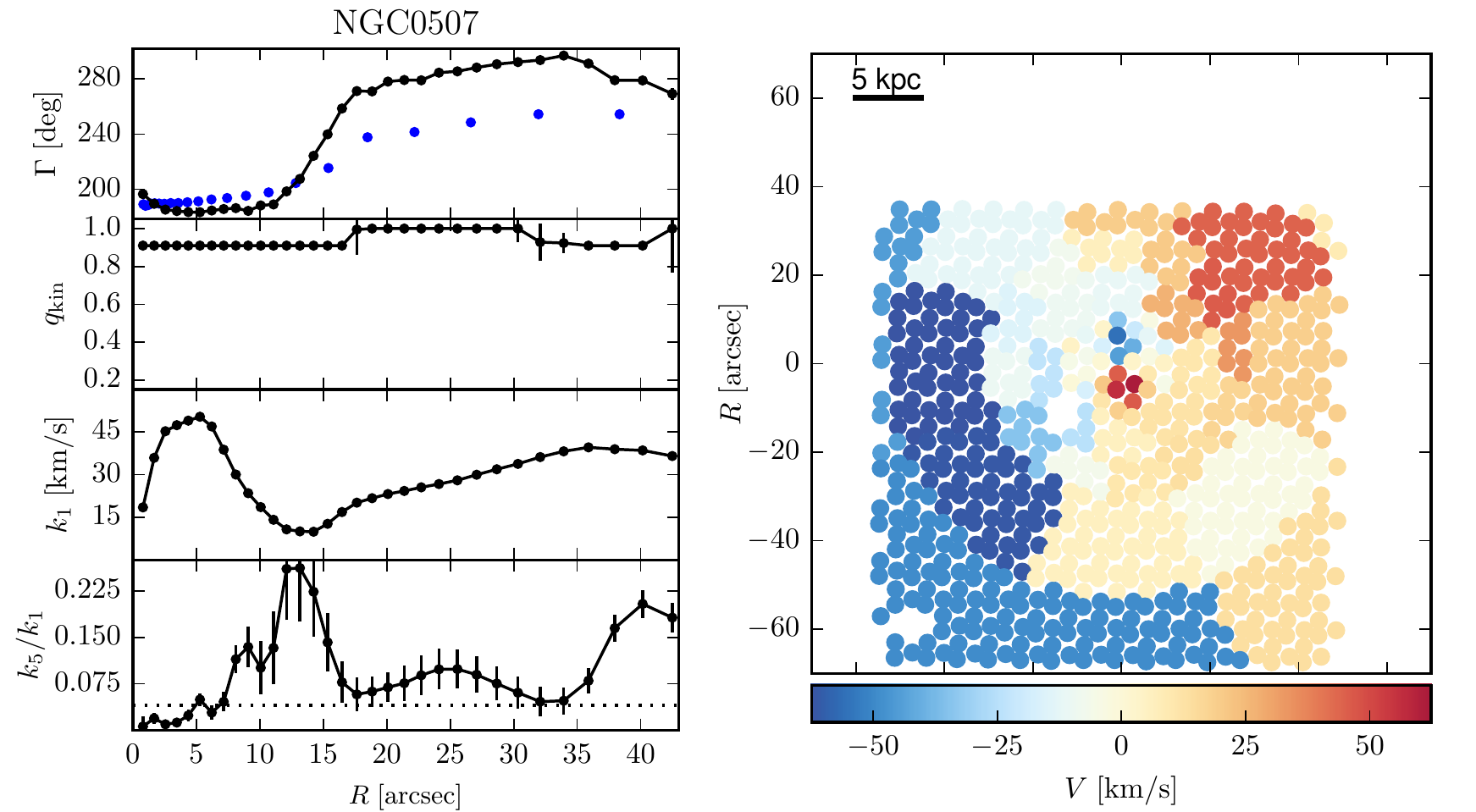}
	\caption{Kinematically distinct components in NGC 507. Left: Radial profiles for coefficients extracted by kinemetry: local position angle $\Gamma$ (measured from North to the receding part), flattening $q_{\rm kin}$, circular velocity amplitude $k_1$ and higher-order deviations $k_5/k_1$ (see Eq.~\ref{eqn:kinemetry}).
		Our HST WFC3 photometric position angle (offset by a constant of $180\degr$) is over-plotted in blue (top panel) and shows a similarly abrupt chance at $R\sim 10$'' to 20'' as the kinematic position angle $\Gamma$.  Right: Velocity map showing the individual Mitchell IFS fibers. Each fiber is colored according to the velocity of the bin it belongs to.}
	\label{fig:0507}
\end{figure*}

\begin{figure*}
	\includegraphics[]{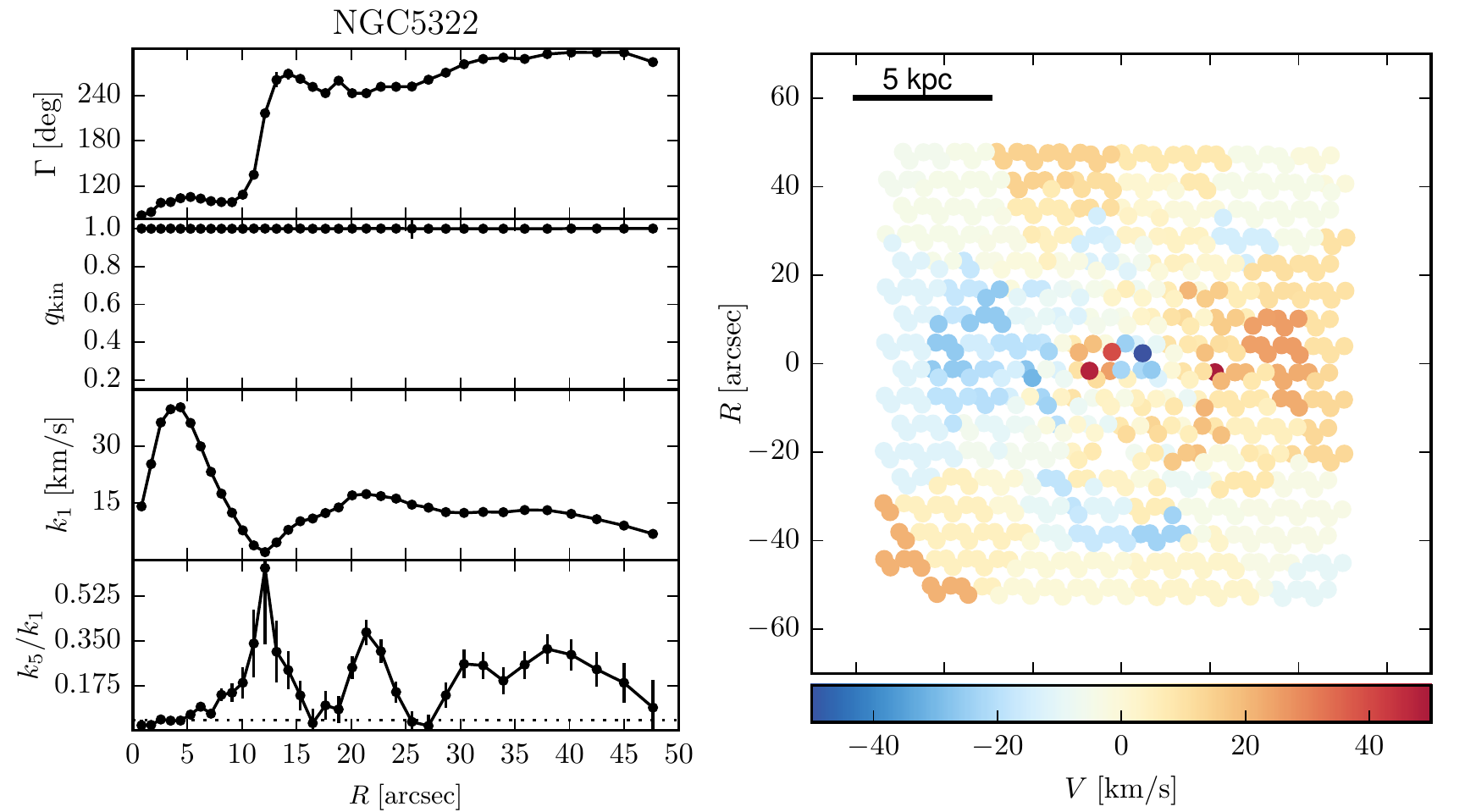}
	\caption{Kinematically distinct components in NGC 5322. The panels show the same quantites as in Fig.~\ref{fig:0507}. The median PA$_\text{kin}$ of the inner and outer kinematic components differ by nearly $180\degr$; that is, the two components are counter-rotating around a similar axis.
	}
	\label{fig:5322}
\end{figure*}

\section{Kinematic versus photometric axis of individual MASSIVE galaxies}

\begin{figure*}
  \includegraphics[width=\textwidth]{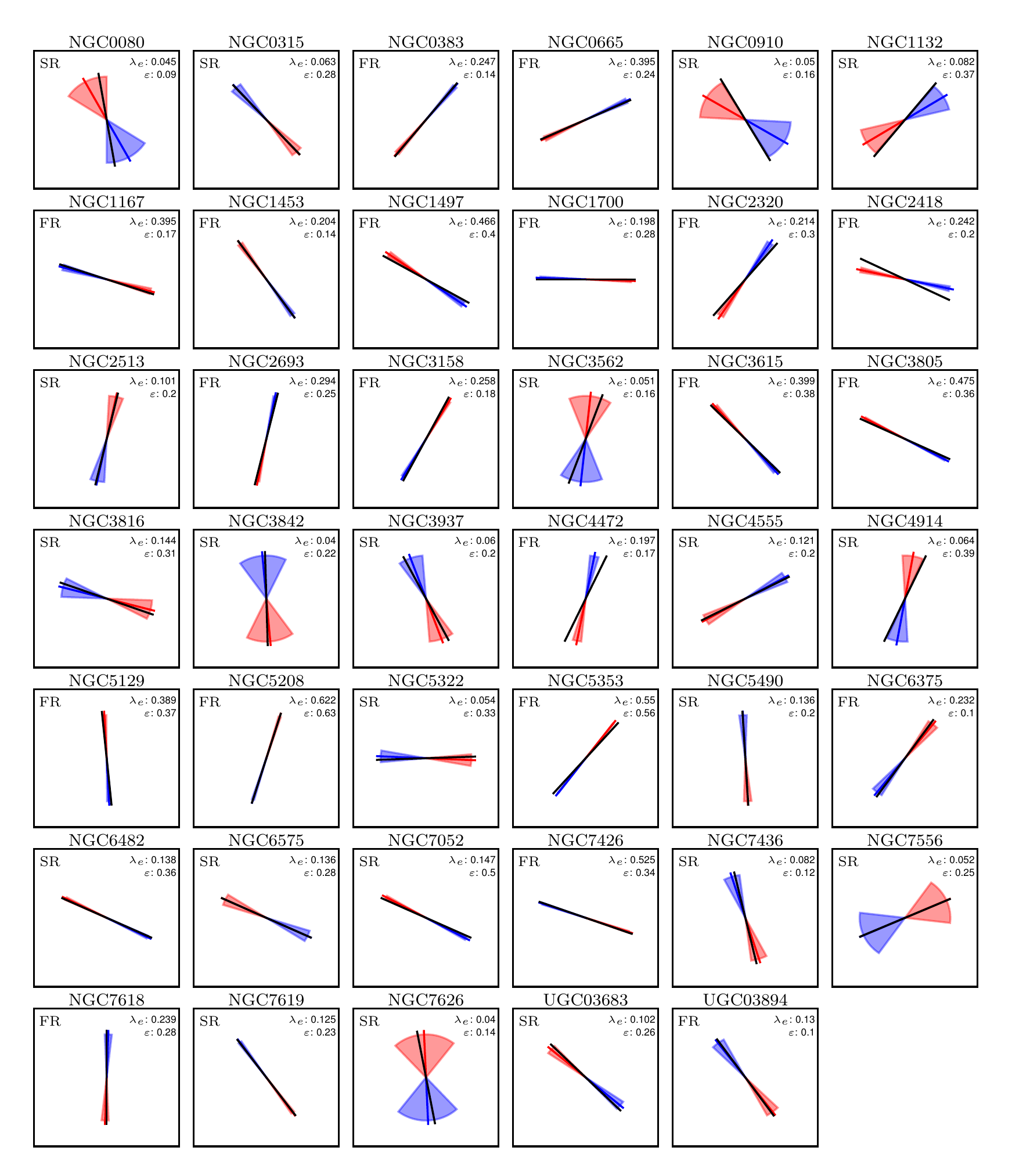}
    \caption{Aligned galaxies. The black line represents the measured photometric PA, PA$_{\text{phot}}$. The colored line and shaded wedge represent the best-fitting kinematic PA, PA$_{\text{kin}}$, and its 1-sigma error -- red indicates the receding part, while blue indicates the approaching part. All plots are oriented with north pointing up and east pointing to the left.}
    \label{fig:aligned}
\end{figure*}

\begin{figure*}
  \includegraphics[width=\textwidth]{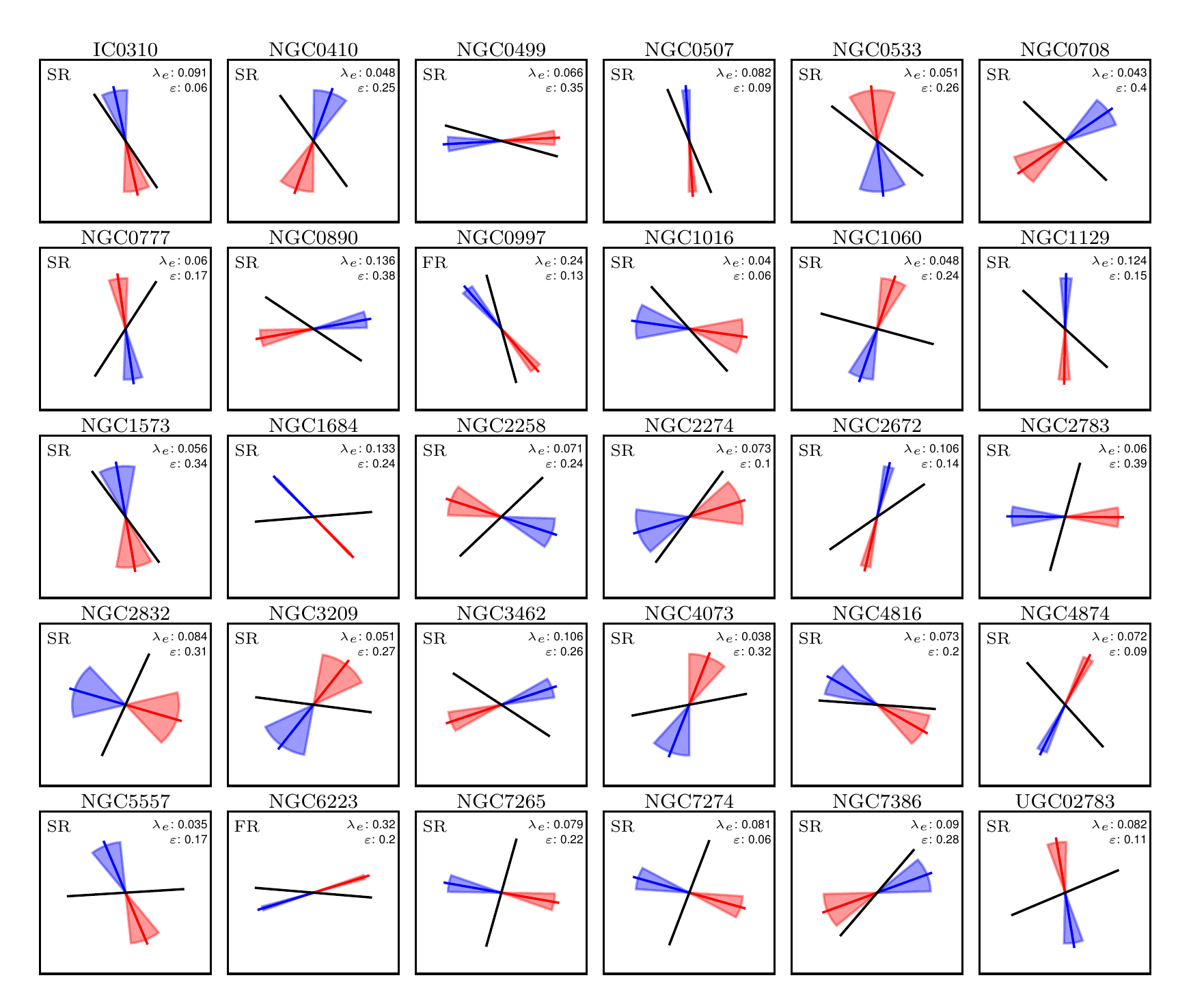}
    \caption{Misaligned galaxies. The black line represents the measured photometric PA, PA$_{\text{phot}}$. The colored line and shaded wedge represent the best-fitting kinematic PA, PA$_{\text{kin}}$, and its 1-sigma error -- red indicates the receding part, while blue indicates the approaching part. All plots are oriented with north pointing up and east pointing to the left.}
    \label{fig:misaligned}
\end{figure*}


\bsp	
\label{lastpage}
\end{document}